\documentclass[aps,prd,superscriptaddress,showpacs,preprintnumbers]{revtex4}
\usepackage{graphicx,color}
\newcommand{\be}{\begin{equation}}
\newcommand{\ee}{\end{equation}}
\newcommand{\bea}{\begin{eqnarray}}
\newcommand{\eea}{\end{eqnarray}}

\newcommand{\bfk}{\mbox{\boldmath $k$}}
\newcommand{\bfh}{\mbox{\boldmath $h$}}
\newcommand{\bfp}{\mbox{\boldmath $p$}}

\newcommand{\bfP}{\mbox{\boldmath $P$}}
\newcommand{\bfS}{\mbox{\boldmath $S$}}
\newcommand{\bfx}{\mbox{\boldmath $x$}}
\newcommand{\bfy}{\mbox{\boldmath $y$}}
\newcommand{\bfz}{\mbox{\boldmath $z$}}
\newcommand{\bfX}{\mbox{\boldmath $X$}}
\newcommand{\bfY}{\mbox{\boldmath $Y$}}
\newcommand{\bfZ}{\mbox{\boldmath $Z$}}

\newcommand{\pup}{p^\uparrow}

\newcommand{\qup}{q^\uparrow}

\newcommand{\la}{\lambda}
\newcommand{\hf}{\hat f}

\def\lsim{\mathrel{\rlap{\lower4pt\hbox{\hskip1pt$\sim$}}\raise1pt\hbox{$<$}}}
\def\gsim{\mathrel{\rlap{\lower4pt\hbox{\hskip1pt$\sim$}}\raise1pt\hbox{$>$}}}
\def\nostrocostruttino#1\over#2{\mathrel{\mathop{\kern 0pt \rlap
{\hbox{$#1$}}} \hbox{\kern-.135em $#2$}}}
\def\sumint{\nostrocostruttino \sum \over {\displaystyle\int}}

\def\kt{k_\perp}
\def\bkt{\bfk_\perp}

\def\pp{p_\perp}
\def\bpp{\bfp_\perp}

\def\xb{x_{_{\!B}}}

\def\avk{\langle k_\perp ^2\rangle}
\def\avp{\langle p_\perp ^2\rangle}
\def\avPT{\langle P_T^2\rangle}
\def\S{_{_S}}
\def\T{_{_T}}
\def\C{_{_C}}
\def\Lsub{_{_L}}
\def\BM{_{_{B\!M}}}

\def\LT{_{_{LT}}}
\def\TL{_{_{TL}}}
\def\TT{_{_{TT}}}
\begin{document}
\preprint{JLAB-THY-11-3}
\title{General Helicity Formalism for Polarized Semi-Inclusive
       Deep Inelastic Scattering}
\author{M.~Anselmino}
\affiliation{Dipartimento di Fisica Teorica, Universit\`a di Torino,
             Via P.~Giuria 1, I-10125 Torino, Italy}
\affiliation{INFN, Sezione di Torino, Via P.~Giuria 1, I-10125 Torino, Italy}
\author{M.~Boglione}
\affiliation{Dipartimento di Fisica Teorica, Universit\`a di Torino,
             Via P.~Giuria 1, I-10125 Torino, Italy}
\affiliation{INFN, Sezione di Torino, Via P.~Giuria 1, I-10125 Torino, Italy}
\author{U.~D'Alesio}
\affiliation{Dipartimento di Fisica, Universit\`a di Cagliari,
             I-09042 Monserrato (CA), Italy}
\affiliation{INFN, Sezione di Cagliari,
             C.P.~170, I-09042 Monserrato (CA), Italy}
\author{S.~Melis}
\affiliation{Dipartimento di Scienze e Tecnologie Avanzate, Universit\`a del Piemonte Orientale, \\
             Viale T.~Michel 11, I-15121 Alessandria, Italy}
\affiliation{INFN, Gruppo Collegato di Alessandria, Via P.~Giuria 1, I-10125 Torino, Italy}
\author{F.~Murgia}
\affiliation{INFN, Sezione di Cagliari,
             C.P.~170, I-09042 Monserrato (CA), Italy}
\author{E.R.~Nocera}
\affiliation{Dipartimento di Fisica Teorica, Universit\`a di Torino,
             Via P.~Giuria 1, I-10125 Torino, Italy}
\author{A.~Prokudin}
\affiliation{Jefferson Laboratory, 12000 Jefferson Avenue, Newport News, VA 23606}
%
%
%

\begin{abstract}
We study polarized Semi-Inclusive Deep Inelastic Scattering (SIDIS) processes,
$\ell(S_\ell) + p(S) \to \ell^\prime \, h \, X$, within the QCD parton model
and a factorization scheme, taking into account all transverse motions, of
partons inside the initial proton and of hadrons inside the fragmenting partons.
We use the helicity formalism. The elementary interactions are
computed at LO with non collinear exact kinematics, which introduces phases
in the expressions of their helicity amplitudes. Several Transverse Momentum
Dependent (TMD) distribution and fragmentation functions appear and contribute to
the cross sections and to spin asymmetries. Our results agree with those
obtained with different formalisms, showing the consistency of our approach.
The full expression for single and double spin asymmetries $A_{S_\ell S}$ is
derived. Simplified, explicit analytical expressions, convenient for
phenomenological studies, are obtained assuming a factorized Gaussian
dependence on intrinsic momenta for the TMDs.

\noindent
\end{abstract}


\maketitle

\section{\label{Intro} Introduction}

Experiments with inclusive Deep Inelastic Scattering (DIS) processes,
$\ell \, N \to \ell' \, X$, have been performed for decades and have
been interpreted as the most common way to investigate the internal
structure of protons and neutrons. At large energy and momentum transfer
the leptons interact with the nucleon constituents; by
detecting the angle and the energy of the scattered lepton one obtains
information on the partonic content of the nucleons. This information is
encoded in the Parton Distribution Functions (PDFs) which give the number
density of partons moving collinearly with the nucleon and carrying a
fraction $x$ of its momentum at a certain value of the squared momentum
transfer $Q^2$. The prediction of the $Q^2$ dependence of the PDFs has been one
of the great successes of pQCD. Although successful, such an approach only
offers information on the longitudinal degrees of freedom of quarks and gluons,
giving no information on the transverse motion, which is integrated over.
This transverse motion -- transverse with respect to the parent nucleon
direction -- is related to intrinsic properties of the partons, like orbital
motion, and reveals new aspects of the nucleon structure.

In the last years, driven by unexpected spin effects and azimuthal dependences,
the study of the intrinsic motion of partons has made enormous progress; indeed,
a new phase in the exploration of the proton and neutron composition has begun.
The leading role in such an effort is played by Semi-Inclusive Deep Inelastic
Scattering (SIDIS) processes, $\ell \, N \to \ell' \, h \, X$, in which, in
addition to the scattered lepton, also a final hadron is detected; this hadron
is generated in the fragmentation of the scattered quark (or gluon) -- 
the so-called current fragmentation region -- and, as such,
yields some new information on the parton primordial motion.
This new information is encoded in the so-called Transverse Momentum Dependent
partonic distribution and fragmentation functions (TMD-PDFs and TMD-FFs, or,
shortly, TMDs), $\hat f_{a/p}(x, \bfk_\perp)$ and $\hat D_{h/a}(z, \bfp_\perp)$.
The TMD-PDFs give the number density of quarks ($a=q$) or gluons ($a=g$) with
light-cone momentum fraction $x$ and transverse momentum $\bfk_\perp$ inside
a fast moving proton; the TMD-FFs give the number density of hadrons $h$
resulting in the fragmentation of parton $a$, with a light-cone momentum
fraction $z$ and a transverse momentum $\bfp_\perp$, relative to the original
parton motion. At leading-twist, taking into account the parton and the nucleon
spins, there are eight independent TMD-PDFs \cite{Mulders:1995dh, Anselmino:2005sh};
if the final hadron is unpolarized or spinless,
say a pion, there are two TMD-FFs. All these quantities combine into physical
observables and by gathering information about them one accesses the momentum
distribution of partons inside the nucleons.

The theoretical framework used to analyze the experimental data is the QCD
factorization scheme, according to which the SIDIS cross section is written
as a convolution of TMDs and elementary interactions:
\be
d\sigma^{\ell p \to \ell' h X} = \sum_q \hat f_{q/p}(x, \bfk_\perp; Q^2)
\otimes d\hat\sigma^{\ell q \to \ell q} \otimes
\hat D_{h/q}(z, \bfp_\perp; Q^2) \>. \label{fac}
\ee
In the $\gamma^* - p$ c.m.~frame, see Fig.~1, the measured transverse momentum,
$\bfP_T$, of the final hadron is generated by the transverse momentum of the
quark in the target proton, $\bfk_\perp$, and of the final hadron with respect
to the fragmenting quark, $\bfp_\perp$. At order $k_\perp/Q$ it is simply given by
\be
\bfP_T =  z \,\bfk_\perp + \bfp_\perp \>.
\ee

There is a general consensus~\cite{Collins:1992kk, Collins:2004nx, Ji:2004wu,
Ji:2004xq, Bacchetta:2008xw} that such a scheme holds in the kinematical
region defined by
\be
P_T \simeq \Lambda_{\rm QCD} \ll Q \>.\label{kin}
\ee
The presence of the two scales, small $P_T$ and large $Q$, allows to
identify the contribution from the unintegrated partonic distribution
($P_T \simeq \kt$), while remaining in the region of validity of the QCD
parton model. At larger values of $P_T$ other mechanisms, like quark-gluon
correlations and higher order pQCD contributions become
important~\cite{Ji:2006br,Anselmino:2006rv,Bacchetta:2008xw}. A
similar situation~\cite{Collins:1984kg,Boer:1999mm,Anselmino:2002pd,
Collins:2004nx,Ji:2004xq,Ji:2006ub,Ji:2006vf,
Arnold:2008kf,Anselmino:2009st} holds for Drell-Yan processes,
$A B \to \ell^+ \ell^- X$, where the two scales are the small transverse
momentum, $q_T$, and the large invariant mass, $M$, of the dilepton pair.

Let us elaborate now on Eq. (\ref{fac}). We consider the SIDIS cross section
at the leading $\alpha_{\rm em}$ order -- {\it i.e.} one-photon exchange --
and in the ``standard" \cite{Bacchetta:2004jz} kinematical configuration of
Fig. 1, which defines the azimuthal angles $\phi_h$ and $\phi_S$ in the
$\gamma^* - p$ c.m. frame. The most general dependence on these angles has been
discussed in several seminal papers \cite{Gourdin:1973qx, Kotzinian:1994dv,
Mulders:1995dh, Bacchetta:2004zf}, both in a
model independent scheme and in the parton model. According to the usual
derivation, the polarization states of the virtual photon, as emitted by the
lepton in a certain direction, contains azimuthal dependences \cite{Gourdin:1973qx,
Kotzinian:1994dv}; within the parton model, the virtual photon scatters off
a quark -- which subsequently fragments into the final hadron -- and each
term of the azimuthal dependences can be written as a convolution of
distribution and fragmentation functions \cite{Kotzinian:1994dv,
Mulders:1995dh, Bacchetta:2004zf, Diehl:2005pc, Bacchetta:2006tn}.
%
\begin{figure}[t]
\begin{center}
\includegraphics[width=14.0truecm]{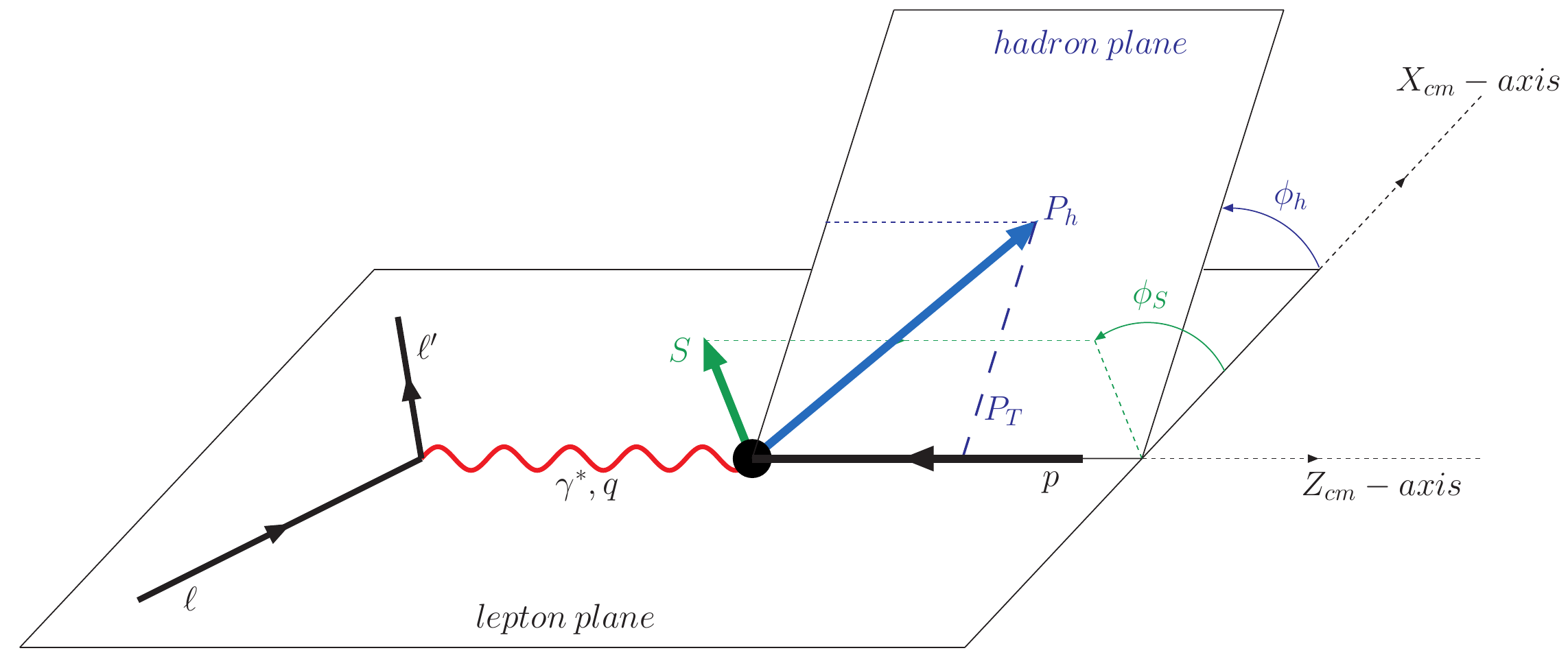}
\caption{\small Kinematical configuration and conventions for SIDIS processes.
The initial and final lepton momenta define the $(X-Z)_{cm}$ plane.}
\label{fig.1}
\end{center}
\end{figure}
%

We re-derive here the same general expression of the cross section, and its
parton model content, by assuming from the beginning the validity of the TMD
factorization (\ref{fac}); we use the helicity basis to compute the elementary
interaction and to introduce transverse momentum dependent distribution and
fragmentation functions. In such an approach the full azimuthal dependence
is simply generated by the properties of the helicity spinors and amplitudes.
Our final results coincide with the existing ones, showing the full
equivalence of the two procedures.
Our formalism is based on a physical and intuitive picture, which somehow
factorizes the physical process in different steps: the ``emission'' of a parton
by the interacting hadron ($p \to q + X$), the interaction of the parton with
the lepton ($\ell \, q \to \ell \, q)$, and the ``emission'' of the final hadron
by the scattered quark ($q \to h +X$); each step is described by the
corresponding helicity amplitudes. For SIDIS processes this factorization has
been formally proven and expressed in terms of TMDs, Eq. (\ref{fac}). Such
a procedure can naturally be extended to other processes, and indeed this has
been done for the large $P_T$ production of a single particle in inclusive
hadronic interactions, $A \, B \to C \, X$ \cite{Anselmino:2005sh}. The point,
however, is that, despite the natural simplicity of the approach, the TMD
factorization has not been proven for processes with a single large scale,
like $A \, B \to C \, X$. Due to this, the study of dijet production at large
$P_T$ in hadronic processes was proposed \cite{Boer:2003tx,Bomhof:2004aw,
Bacchetta:2005rm, Bomhof:2007su}, where the second small scale is the total
$q_T$ of the two jets, which is of the order of the intrinsic partonic momentum
$k_\perp$. This procedure leads to a modified TMD factorization approach,
with the inclusion in the elementary processes of gauge link color factors
\cite{Bomhof:2006dp,Bacchetta:2007sz,Ratcliffe:2007ye,Gamberg:2010tj}.
However, some doubts on the validity of such a factorization scheme have been
recently cast \cite{Rogers:2010dm}. A possible experimental test of the TMD
factorization for processes with only one large scale has been proposed in
Ref.~\cite{Anselmino:2009pn}.
We limit our discussion in this paper to SIDIS processes, in the kinematical
region (\ref{kin}) for which TMD factorization holds, and obtain the most general expression for the polarized cross section, with our helicity formalism.
A similar study can be done, with the same validity, for Drell-Yan processes
\cite{Anselmino:2002pd,Anselmino:2009st,Anselmino:2009zza}. We introduce only
leading-twist TMDs and take into account exact kinematics, often simplifying
results by only keeping terms up to ${\cal O}\left( k_\perp /Q \right)$.

The paper is organized as follows. In Section II we present our formalism and
compute the polarized SIDIS cross section. In Section III we give the explicit
general expressions of all independent single and double spin asymmetries, in
terms of the TMDs. In Section IV we give explicit analytical formulae for the
spin and azimuthal asymmetries, assuming a factorized Gaussian dependence of
the TMDs on $k_\perp$ and $p_\perp$. In Section V we draw our conclusions.
Useful results are derived and collected in Appendices A--E.

\section{\label{XS} Cross sections in polarized SIDIS}

According to Refs. \cite{Anselmino:2005nn} and \cite{Anselmino:2005sh}
the full differential cross section for the polarized SIDIS process,
$\ell(S_\ell) \, + \, p(S)  \to \ell^\prime \, h \, X$, can be written,
within TMD factorization, as
\bea
\frac{d\sigma^{\ell (S_\ell) + p (S) \to \ell^\prime h X}}
{d\xb dQ^2 \, dz_h\, d^2 \bfP _T \, d\phi_S} &=&
\frac {1}{2\pi} \sum_{q} \sum_{\{\lambda\}} \>
\frac {1}{16 \, \pi \, (\xb s)^2}\int d^2\bkt \, 
\frac {z}{z_h} \, J \; 
\nonumber \\
&\times& \rho^{\ell,S_\ell}_{\lambda^{\,}_\ell \lambda^\prime_\ell} \,
\rho^{q_i/p,S}_{\lambda^{\,}_{q_i} \lambda_{q_i}^\prime} \,
\hf_{q_i/p,S}(x,\bkt) \,
\hat M^{\,}_{\lambda^{\,}_{\ell^\prime}\lambda^{\,}_{q_f};
\lambda^{\,}_{\ell}\lambda^{\,}_{q_i}} \,
\hat M^*_{\lambda^{\,}_{\ell^\prime}\lambda_{q_f}^\prime;
     \lambda_{\ell}^\prime \lambda_{q_i}^\prime} \,
\hat D^{\la^{\,}_h, \la^{\,}_h} _{\lambda^{\,}_{q_f},
\lambda_{q_f}^\prime}(z,\bfp_\perp) \label{gen-xs}\>
\>,
\eea
where we adopt the kinematical configuration of Fig. 1, and, as usual:
\be
s = (\ell + p)^2 \quad\quad Q^2=-q^2 = -(\ell -\ell')^2 \quad\quad
\xb = \frac {Q^2}{2p \cdot q}\quad\quad
z_h = \frac{p \cdot P_h}{p \cdot q} \> \cdot
\ee
The variables $x$, $z$ and $\bfp _\perp$ which appear under integration in
Eq. (\ref{gen-xs}) are related to the final observed variables $\xb$, $z_h$
and $\bfP_T$ and to the integration variable $\bfk_\perp$. The exact
relations can be found in Ref. \cite{Anselmino:2005nn};
at ${\cal O}(k_\perp/Q)$ one simply has
\be
x = \xb \quad\quad z = z_h \quad\quad
\bfp_\perp = \bfP_T - z_h \bfk_\perp \>.
\ee
$J$ includes some non-planar kinematical factors \cite{Anselmino:2005nn}:
\be
J = \frac{\xb}{x}\left( 1 + \frac{\xb^2}{x^2}\frac{\kt^2}{Q^2}
\right)^{\!\!-1} \simeq 1 \,,
\ee
where the last relation holds at ${\cal O}(k_\perp/Q)$. At this order
Eq. (\ref{gen-xs}) can be written as:
\bea
\frac{d\sigma^{\ell(S_\ell) + p(S) \to \ell^\prime h X}}
{d\xb \, dQ^2 \, dz_h \, d^2 \bfP _T \, d\phi_S}
&\simeq& \frac {1}{2\pi} \sum_{q} \sum_{\{\lambda\}}
\> \frac {1}{16 \, \pi \, (\xb s)^2} \int d^2\bkt \, d^2\bfp_\perp
\> \delta ^{(2)}(\bfP_T - z_h\bfk_\perp -\bfp_\perp)
\nonumber \\
&\times& \rho^{\ell, S_\ell}_{\lambda^{\,}_\ell \lambda^\prime_\ell} \,
\rho^{q/p,S}_{\lambda^{\,}_{q_i} \lambda_{q_i}^\prime} \,
\hf_{q_i/p,S}(x,\bkt) \,
\hat M^{\,}_{\lambda^{\,}_{\ell^\prime}\lambda^{\,}_{q_f};
\lambda^{\,}_{\ell}\lambda^{\,}_{q_i}} \,
\hat M^*_{\lambda^{\,}_{\ell^\prime}\lambda_{q_f}^\prime;
     \lambda_{\ell}^\prime \lambda_{q_i}^\prime} \,
\hat D^{\la^{\,}_h, \la^{\,}_h}
_{\lambda^{\,}_{q_f}, \lambda_{q_f}^\prime}(z,\bfp_\perp) \>
\>, \label{app-xs}
\eea
where we have explicitly shown the integration over $\bfp _\perp$ for clarity 
and further use. In Eqs. (\ref{gen-xs}) and (\ref{app-xs}) the sums are 
performed over all quark flavors ($q = u, \bar u, d, \bar d, s, \bar s$) and 
all quark, lepton and hadron helicity indices; 
$\rho^{\ell, S_\ell}_{\lambda^{\,}_\ell \lambda^\prime_\ell}$
is the initial lepton helicity density matrix, which describes the spin state 
of the lepton beam; for unpolarized leptons one simply has
$\rho^{\ell}_{\lambda^{\,}_\ell \lambda^\prime_\ell} =
\frac 12 \, \delta_{\lambda^{\,}_\ell \lambda^\prime_\ell}$. It might be
helpful, and useful for physical interpretations, to recall that, in general,
for a spin 1/2 Dirac particle one has:
\be
\rho^{\,}_{\la^{\,} \la^{\prime}} =
\frac{1}{2}\,{\left(
\begin{array}{cc}
1+P_{z} & P_{x} - i P_{y} \\
P_{x} + i P_{y} & 1-P_{z}
\end{array} \right)}\,, \label{rho-general}
\ee
where $P_j = P_{x}, P_{y}, P_{z}$ are the components of the particle
polarization vector in its helicity frame
(throughout the paper we follow the definitions and conventions for helicity
states of Ref.~\cite{Leader:2001gr}).

Let us discuss in detail the different ``factors" in Eq. (\ref{gen-xs}): they represent the
distribution of polarized partons (only quarks at LO) inside the proton, their
interaction with the lepton and the fragmentation of the (polarized) final quark
into the observed unpolarized hadron $h$. We follow, and adapt to the case of SIDIS, the
discussion of Ref.~\cite{Anselmino:2005sh}. We describe the three stages of the
process -- quark emission, interaction and fragmentation -- within the helicity
formalism, which allows us to introduce in a natural way, at each step, several
phases; these, when combined into the expression for the physical cross section
(\ref{gen-xs}) give its full azimuthal dependence, in agreement with results
in the literature derived in a more formal and somewhat less intuitive
way \cite{Bacchetta:2006tn}.

\subsection{TMD partonic distribution functions \label{soft-pdf}}

$\rho^{q_i/p,S}_{\lambda^{\,}_{q_i} \lambda_{q_i}^\prime} \, \hf_{q_i/p,S}(x,\bkt)$
counts the number of polarized quarks inside a polarized proton; it is the
polarized distribution function of the initial quark $q_i$ with
light-cone momentum fraction $x$ and intrinsic transverse momentum $\bkt$,
inside the target proton $p$ in a pure spin state $S$. Using Eq.~(\ref{rho-general}) and
parity invariance one can see that there are eight independent distribution
functions, which can be defined as:
\bea
&&P^{q}_j \, \hf _{q/p,S_T}(x, \bfk_{\perp}) =
\hf _{s_j/S_T}^{q}(x, \bkt) -  \hf _{-s_j/S_T}^{q}(x, \bkt) \equiv
\Delta \hf_{s_j/S_T}^{q}(x, \bfk_{\perp}) \label{DxY}\\
&&P^{q}_j \, \hf _{q/p,S_L}(x, \bfk_{\perp}) =
\hf _{s_j/S_L}^{q}(x, \bkt) -  \hf _{-s_j/S_L}^{q}(x, \bkt) \equiv
\Delta \hf_{s_j/S_L}^{q}(x, \bfk_{\perp}) \label{DxZ}\\
&&\hf _{{q}/p,S_T}(x, \bfk_{\perp}) \equiv f_{{q}/p}(x, k_{\perp}) +
\frac{1}{2}\,\Delta \hf_{{q}/S_T}(x, \bfk_{\perp})\,,\label{Dunp}
\label{main-table}
\eea
with
\be
\Delta \hf_{{q}/S_T}(x, \bfk_{\perp}) \equiv  \hf _{q/S_T}(x, \bkt) -  \hf _{q/-S_T}(x, \bkt) \,.
\ee
We define, for further use,
\be
\frac{1}{2}\,[\hf _{s_y/S_T}(x, \bkt) -  \hf _{s_y/-S_T}(x, \bkt)] \equiv \Delta ^- \hf_{{s_y}/S_T}(x, \bfk_{\perp}).
\ee
In Eqs.~(\ref{DxY}) and (\ref{DxZ}), $j = {x}, {y}, {z}$ are the coordinate-axes in the quark helicity frame
and $S_{L,T}$ are respectively the longitudinal and transverse components of
the proton polarization vector, with respect to its direction of motion.

Different notations can be found in the literature for these functions,
in particular those introduced by the Amsterdam
group~\cite{Mulders:1995dh,Boer:1997nt,Boer:2003cm}, which are largely
adopted. The relationships between the two sets can be found in
Ref.~\cite{Anselmino:2005sh}, and will be repeated for convenience in
Eqs.~(\ref{fqs})--(\ref{fyqs}).

According to the physical interpretation of the factorization scheme,
as outlined above, these quantities can be introduced by making use of the
helicity amplitudes
${\hat{\cal F}}_{\la^{\,}_q, \la^{\,}_{X}; \la^{\,}_p}$, which describe the
soft process $p \to q + X$. Since the partonic distribution
is usually regarded, at LO, as the inclusive cross section for this process,
the helicity density matrix of a quark $q$ inside the proton $p$ with spin $S$
can be written as
\bea
\rho_{\la^{\,}_q \la^{\prime}_q}^{q/p,S} \>
\hat f_{q/p,S}(x,\bfk_{\perp })
&=& \sum _{\la^{\,}_p, \la^{\prime}_p}
\rho_{\la^{\,}_p \la^{\prime}_p}^{p,S}
\sumint_{X, \la_{X}} \!\!\!\!\!
{\hat{\cal F}}_{\la^{\,}_q, \la^{\,}_{X};
\la^{\,}_p} \, {\hat{\cal F}}^*_{\la^{\prime}_q,\la^{\,}_{X}; \la^{\prime}_p}
\nonumber \\
&\equiv& \sum _{\la^{\,}_p, \la^{\prime}_p}
\rho_{\la^{\,}_p \la^{\prime}_p}^{p,S} \>
\hat F_{\la^{\,}_p, \la^{\prime}_p}^{\la^{\,}_q,\la^{\prime}_q} \>,\label{defF}
\eea
having defined
\be
\hat{F}_{\la^{\,}_p, \la^{\prime}_p}^{\la^{\,}_q,\la^{\prime}_q} \equiv \>
\sumint_{X, \la_{X}} \!\!\!\!\!\!
{\hat{\cal F}}_{\la^{\,}_q,\la^{\,}_{X};\la^{\,}_p} \,
{\hat{\cal F}}^*_{\la^{\prime}_q,\la^{\,}_{X}; \la^{\prime}_p} \>,
\label{defFF}
\ee
where the $\sumint_{X, \la_{X}}\!\!\!$ stands for a spin sum and phase-space integration
over all the undetected remnants of the proton,
considered as a system $X$, and the $\hat{\cal F}$'s are the
{\it helicity distribution amplitudes} for the $p \to q + X$ process.
Eq.~(\ref{defF}) relates, via the unknown distribution amplitudes,
the helicity density matrix of the parton $q$,
\be
\rho_{\la^{\,}_q \la^{\prime}_q}^{q/p,S} =
\frac{1}{2}\,{\left(
\begin{array}{cc}
1 + P_z^q & P_x^q - i P_y^q \\
P_x^q + i P_y^q & 1 - P_z^q
\end{array}
\right)} = \frac{1}{2}\,{\left(
\begin{array}{cc}
1 + P_z^q  & P_T^q \, e^{-i \varphi_{s_q}}  \\
P_T^q \, e^{i\varphi_{s_q}} & 1 - P_z^q
\end{array}
\right)} \,, \label{rho-q}
\ee
to the helicity density matrix of the polarized parent proton,
\be
\rho_{\la^{\,}_p \la^{\prime}_p}^{p,S} =
\frac{1}{2}\,{\left(
\begin{array}{cc}
1 + S_Z & S_X - i S_Y \\
S_X + i S_Y & 1 - S_Z
\end{array}
\right)} = \frac{1}{2}\,{\left(
\begin{array}{cc}
1 + S_L & S_T \, e^{-i\varphi_{S}}  \\
S_T \, e^{i\varphi_{S}} & 1 - S_L
\end{array}
\right)} \,. \label{rho-p}
\ee
In the above equations $\bfS = (S_X, S_Y, S_Z) = (S_T \cos \varphi_{S}, S_T
\sin \varphi_{S}, S_L)$ is the proton polarization vector and $\varphi_{S}$ its
azimuthal angle, defined in the helicity reference frame of the proton $p$.
Similarly, $\bfP ^q = (P_x^q, P_y^q, P_z^q) = (P^q_T \cos \varphi_{s_q},
P^q_T \sin \varphi_{s_q}, P^q_z)$ is the quark polarization vector defined in
the quark helicity frame and $\varphi_{s_q}$ its azimuthal angle.
For the kinematical configuration of Fig.~1, one has $\varphi_S = 2 \pi - \phi_S$
(see Appendix \ref{hel-transf}), so that:
\be
\rho_{\la^{\,}_p \la^{\prime}_p}^{p,S} = \frac{1}{2}\,{\left(
\begin{array}{cc}
1 + S_L & S_T \, e^{i\phi_{S}}  \\
S_T \, e^{-i\phi_{S}} & 1 - S_L
\end{array}
\right)}\,. \label{rho-p-cm}
\ee
Notice that, in general, we denote by $\varphi$ angles defined in the proton
or quark helicity frames, while the symbol $\phi$ is used for the corresponding angles measured
in the $\gamma^*-p$ c.m. frame.

The distribution amplitudes $\hat{\cal F}$ depend on the parton light-cone
momentum fraction $x$ and on its intrinsic transverse momentum
$\bfk_{\perp}$, with modulus $k_{\perp }$ and azimuthal angle $\phi_\perp$,
in a precise way \cite{Anselmino:2005sh,Leader:2001gr}, which, again referred
to the kinematical configuration of Fig.~1, reads:
\be
\hat{\cal F}_{\la^{\,}_q,\la^{\,}_{X}; \la^{\,}_p}(x, \bfk_{\perp})
={\cal F}_{\la^{\,}_q,\la^{\,}_{X}; \la^{\,}_p}(x, k_{\perp}) \>
{\rm exp}[-i\la^{\,}_p \phi_\perp] \>,
\label{dampphi}
\ee
so that
\be
\hat F_{\la^{\,}_p,\la^{\prime}_p}^{\la^{\,}_q,\la^{\prime}_q}(x,\bfk_{\perp})
= F_{\la^{\,}_p,\la^{\prime}_p}^{\la^{\,}_q,\la^{\prime}_q}(x, k_{\perp})
\> {\rm exp}[i(\la^{\prime}_p - \la^{\,}_p)\phi_\perp] \>. \label{fft-ff}
\ee
$F_{\la^{\,}_p,\la^{\prime}_p}^{\la^{\,}_q,\la^{\prime}_q}(x, k_{\perp})$
has the same definition as
$\hat{F}_{\la^{\,}_p,\la^{\prime}_p}^{\la^{\,}_q,\la^{\prime}_q}
(x, \bfk_{\perp})$, Eq.~(\ref{defFF}), with $\hat{\cal F}$ replaced by
${\cal F}$, and does not depend on phases anymore. Notice that we have chosen,
throughout the paper, to denote with a hat  all soft quantities
which depend on both the modulus and the phase of the $\bkt$ and $\bpp$ intrinsic
momentum vectors, while we drop the hat for quantities which only
depend on the modulus of these vectors and not on their phases.

Eqs. (\ref{defF}), (\ref{rho-q}), (\ref{rho-p-cm}) and (\ref{fft-ff}), together
with parity properties and the arguments collected in Appendix B, allow to extract
the explicit phase dependence of the eight polarized distribution
functions~(\ref{DxY})--(\ref{Dunp}), with the result
(more details can be found in Ref.~\cite{Anselmino:2005sh}):
\bea
\hat f_{q/p,S} (x,\bkt) &=&
F^{++}_{++} (x,\kt) + F^{++}_{--} (x,\kt) -
2 \, S_T \,{\rm Im} F^{++}_{+-}(x,\kt) \, \sin (\phi_S-\phi_\perp) \label{fqs}\\
&=& f_{q/p}(x,\kt) - \frac{1}{2} \, S_T \,\Delta f_{q/S_T}(x,\kt)
\sin (\phi_S-\phi_\perp) \nonumber \\
&=& f_1(x,\kt) +  S_T \, \frac{\kt}{M} \, f_{1T}^{\perp}(x,\kt)
\, \sin (\phi_S-\phi_\perp) \nonumber \\
P_z^{q} \, \hat f_{q/p,S} (x,\bkt)&=&
S_L \left[ F^{++}_{++} (x,\kt)-F^{++}_{--} (x,\kt) \right] +
2 \, S_T \,{\rm Re} F^{++}_{+-} (x,\kt) \, \cos(\phi_S-\phi_\perp)
\label{fzqs} \\
&=& S_L \,\Delta f^{q}_{s_z/S_L}(x,\kt) +
S_T \, \Delta f^{q}_{s_z/S_T}(x,\kt) \, \cos(\phi_S-\phi_\perp)\nonumber \\
&=& S_L \, g_{1L}(x,\kt) +  S_T \, \frac{\kt}{M} \, g_{1T}^{\perp}(x,\kt)
\, \cos(\phi_S-\phi_\perp) \nonumber \\
P_x^{q} \, \hat f_{q/p,S} (x,\bkt)&=&
- 2 \, S_L \,{\rm Re} F^{+-}_{++} (x,\kt)
- S_T \left[ F^{+-}_{+-} (x,\kt) + F^{-+}_{+-} (x,\kt) \right] \,
\cos(\phi_S-\phi_\perp) \label{fxqs} \\
&=& - S_L \,\Delta f^{q}_{s_x/S_L}(x,\kt) - S_T \,
\Delta f^{q}_{s_x/S_T}(x,\kt) \, \cos(\phi_S-\phi_\perp)\nonumber \\
&=& - S_L \, \frac{\kt}{M} \, h_{1L}^\perp(x,\kt) -
S_T \left[ h_{1}(x,\kt) + \frac{\kt ^2}{2M^2} \, h_{1T}^{\perp}(x,\kt)\right]
\, \cos(\phi_S-\phi_\perp) \nonumber \\
P_y^{q} \, \hat f_{q/p,S} (x,\bkt)&=&
2 \,{\rm Im} F^{+-}_{++} (x,\kt)
+ S_T \left[ F^{+-}_{+-} (x,\kt) - F^{-+}_{+-} (x,\kt) \right] \,
\sin(\phi_S-\phi_\perp)\label{fyqs} \\
&=& - \Delta f^{q}_{s_y/p}(x,\kt) + S_T \, \Delta ^- f^{q}_{s_y/S_T}(x,\kt)
\, \sin(\phi_{S}-\phi_\perp)\nonumber \\
&=& \frac{\kt}{M} \,  h_{1}^\perp(x,\kt) +  S_T \,
\left[ h_{1}(x,\kt) - \frac{\kt ^2}{2M^2} \, h_{1T}^{\perp}(x,\kt)\right]
\, \sin(\phi_S-\phi_\perp)\,. \nonumber
\eea
As already stated, $\phi_S$ and $\phi_\perp$ are respectively the azimuthal
angle of the proton polarization vector $\bfS$ and of the quark intrinsic
momentum $\bkt$ measured in the $\gamma^* - p$ c.m. frame of Fig.~1. Also
the quark polarization vector components $P_i^q \> (i = x,y,z)$ refer to
the helicity frame of the quark, as reached from the $\gamma^* - p$ frame:
this explains the sign differences between Eqs.~(\ref{fqs}, \ref{fxqs}--\ref{fyqs}) and
Eqs.~(B12, B14--B15) of Ref.~\cite{Anselmino:2005sh} (in the latter case the
polarized proton was moving along $Z_{cm}$ rather than $-Z_{cm}$. Further
comments are given in Appendix B). Notice that, while $P_y^q \, f_{q/p}
\not= 0$, one has $P_x^q \, f_{q/p} = P_z^q \, f_{q/p} = 0$.

The above equations, which will be soon used, deserve some further explanation.
In each equation the first line expresses the partonic distributions in terms
of the $F_{\la^{\,}_p,\la^{\prime}_p}^{\la^{\,}_q,\la^{\prime}_q}(x, k_{\perp})$'s
and shows their exact phase dependence. The second line gives the same
quantities using our notations for the TMD-PDFs. According to our ``hat convention'', quantities
like $\Delta f^{q}_{s_j/S}(x,\kt)$ do not depend on phases anymore, as such
dependence has been explicitly extracted out; comparing with Eqs.~(\ref{DxY})--(\ref{Dunp}) one has (always referred to the variables and kinematical
configuration of Fig.~1):
\bea
\Delta \hf_{{q}/S_T}(x, \bkt) &=& - \Delta f_{{q}/S_T}(x, \kt)
\, \sin(\phi_{S} -\phi_\perp) \label{DD1} \\
\Delta \hf_{s_x/S_L}^{q}(x, \bkt) &=& - \Delta f_{s_x/S_L}^{q}(x, \kt) \label{DD2} \\
\Delta \hf_{s_x/S_T}^{q}(x, \bkt) &=& - \Delta f_{s_x/S_T}^{q}(x, \kt)
\, \cos(\phi_{S} -\phi_\perp)\label{DD3} \\
\Delta \hf_{s_y/S_L}^{q}(x, \bkt) &=& -\Delta f_{s_y/S_L}^{q}(x, \kt)
= -\Delta f_{s_y/p}^{q}(x, \kt) \label{DD4} \\
\Delta \hf_{s_y/S_T}^{q}(x, \bkt) &=&  -\Delta f_{s_y/p}^{q}(x, \kt)
+ \Delta^- f_{s_y/S_T}^{q}(x, \kt) \, \sin(\phi_{S} -\phi_\perp) \label{DD5} \\
&\equiv& -\Delta f_{s_y/p}^{q}(x, \kt) + \Delta^- \hat f_{s_y/S_T}^{q}(x, \bkt)
\nonumber \\
\Delta \hf_{s_z/S_L}^{q}(x, \bkt) &=& \Delta f_{s_z/S_L}^{q}(x, \kt) \label{DD6} \\
\Delta \hf_{s_z/S_T}^{q}(x, \bkt) &=& \Delta f_{s_z/S_T}^{q}(x, \kt)
\, \cos(\phi_{S} -\phi_\perp) \label{DD7} \>.
\eea
According to our choice the $\Delta f_{s_j/S_T,S_L}^{q}(x, \kt)$ introduced
here are the same as in Ref.~\cite{Anselmino:2005sh}.

The last line of Eqs.~(\ref{fqs})--(\ref{fyqs}) gives the connection with the
Amsterdam group notations; $M$ is taken as the proton mass. These last
relationships hold at leading twist; notice also that, when comparing
with the results of the Amsterdam group, one should take into account other
differences in conventions and notations. In particular:
\bea
(\bfp _T)_{\rm Amsterdam} &=& \bfk _\perp  \label{a-noi1}\\
(-z \, \bfk _T)_{\rm Amsterdam}  &=& \bfp _\perp = (\bfP _T - z_h \, \bfk _\perp )
\label{a-noi2} \\
(\hat{\bfh})_{\rm Amsterdam} &=& \frac{\bfP_T}{P_T} = \hat{\bfP}_T \>.
\label{a-noi3}
\eea

Finally, we recall some other notations widely used in the literature:
\be
\Delta^N \! f_{q/\pup}(x, \kt) \equiv \Delta f_{q/S_T}(x,\kt) =
4 \,{\rm Im} F^{++}_{+-}(x,\kt) = - \frac {2\kt}{M} \, f_{1T}^\perp (x, \kt)  \label{siv}
\ee
\vskip -12pt
\be
\Delta^N \! f_{\qup/p}(x, \kt) \equiv \Delta f^{q}_{s_y/p}(x,\kt)
= - 2 \,{\rm Im} F^{+-}_{++}(x,\kt)
= - \frac {\kt}{M} \, h_{1}^\perp (x, \kt) \label{b-m}
\ee
\vskip -12pt
\bea
\frac{1}{2}\left[ \Delta f^q_{s_x/S_T}(x,\kt) +
\Delta^-f^q_{s_y/S_T}(x,\kt) \right]
&=& F^{+-}_{+-}(x,\kt) = h_{1T}(x,\kt)+\frac{\kt ^2}{2M^2}h^\perp_{1T}(x,\kt) \equiv h_1(x,\kt) \label{F+-+-} \\
\frac{1}{2}\left[ \Delta f^q_{s_x/S_T}(x,\kt) -
\Delta^-f^q_{s_y/S_T}(x,\kt)\right] &=& F^{-+}_{+-}(x,\kt) =
\frac{\kt ^2}{2M^2}h^\perp_{1T}(x,\kt) \label{F-++-}
\eea
\vskip -12pt
\be
\Delta _T q(x) =  h_1(x) =
\int \!d^2\bkt \, h_{1}(x,\kt) =
\int \!d^2\bkt
\left[h_{1T}(x,\kt)+\frac{\kt ^2}{2M^2}h^\perp_{1T}(x,\kt)\right]
\label{int-transv} \,.
\ee
Eqs.~(\ref{siv}), (\ref{b-m}) and (\ref{int-transv}) refer, respectively, to the Sivers, the Boer-Mulders and the
transversity distributions.

\subsection{TMD fragmentation functions}

The quantity $\hat D^{\la^{\,}_h, \la^{\prime}_h}_{\lambda^{\,}_{q_f},
\lambda_{q_f}^\prime}(z,\bfp_\perp)$ describes the hadronization of the quark
$q_f$ into the observed final hadron $h$, which carries, with respect to the
fragmenting quark, the light-cone momentum fraction $z$ and the intrinsic
transverse momentum $\bpp$. Similarly to the distribution functions, also
$\hat D^{\la^{\,}_h,\la^{\prime}_h}_{\la^{\,}_q,\la^{\prime}_q}(z,\bfp_{\perp})$
can be written as the product of {\it fragmentation amplitudes} for the
$q \to h + X$ process:
\be
\hat D^{\la^{\,}_h, \la^{\prime}_h}_{\la^{\,}_q, \la^{\prime}_q}
= \> \sumint_{X, \la_{X}} {\hat{\cal D}}_{\la^{\,}_{h},\,\la^{}_X;
\la^{\,}_q} \, {\hat{\cal D}}^*_{\la^{\prime}_h,\,\la^{}_{X}; \la^{\prime}_q}
\, ,
\ee
where the $\sumint_{X, \la_{X}}$ stands for a spin sum and phase space
integration over all undetected particles, considered as a system $X$.
The usual unpolarized fragmentation function  $D_{h/q}(z)$, {\it i.e.} the
number density of hadrons $h$ resulting from the fragmentation of an
unpolarized parton $q$ and carrying a light-cone momentum fraction $z$,
is given by
\be
D_{h/q}(z) = \frac{1}{2} \sum_{\la^{\,}_q,\la^{\,}_h} \int d^2\bfp_{\perp}
\, \hat D^{\la^{\,}_h, \la^{\,}_h}_{\la^{\,}_q, \la^{\,}_q}(z, \bfp_{\perp})
\,. \label{fr}
\ee

We consider only the cases in which the final particle is either spinless
($\la^{\,}_h = 0$) or its polarization is not observed,
\be
D^{h/q}_{\la^{\,}_q, \la^{\prime}_q}(z, \bfp_{\perp}) = \sum_{\la^{\,}_h}
\hat D^{\la^{\,}_h, \la^{\,}_h}_{\la^{\,}_q, \la^{\prime}_q}(z, \bfp_{\perp})
\,. \label{fr2}
\ee
In such a case, parity invariance reduces to two the number of independent
$\hat D^{h/q}_{\la^{\,}_q,\la^{\prime}_q}(z,\bfp_{\perp})$. These, in general,
may depend on the azimuthal angle of the final hadron momentum $\bfP_h$
around the direction of the fragmenting quark $q$, {\it as defined in the
quark $q$ helicity frame}, which we denote by $\varphi_q^{h}$ (it was actually
denoted as $\phi_q^{h}$ in Ref.~\cite{Anselmino:2005sh}):
\bea
&&\hat D_{++}^{h/q}(z,\bpp)=\hat D_{--}^{h/q}(z,\bpp)= D_{h/q}(z,p_\perp)
\label{d++}\\
&&\hat D_{+-}^{h/q}(z,\bpp)= D_{+-}^{h/q}(z,p_\perp)\; e^{i\varphi_q^h}
\label{d+-}\\
&&\hat D_{-+}^{h/q}(z,\bpp)= [D_{+-}^{h/q}(z,\bpp)]^*
= -D_{+-}^{h/q}(z,p_\perp)\; e^{-i\varphi_q^h } \>. \label{d-+}
\eea
In Appendix~\ref{phi} it is shown how to express $\varphi_{q}^{h}$
in terms of integration and external variables (defined in the $\gamma^*-p$
c.m. frame), with the result, at leading order in the $(\kt/Q)$ expansion:
\bea
&& \cos \varphi_q^h =\frac{P_T}{\pp}
         \left[\cos(\phi _h-\phi_\perp) -z_h\frac{\kt}{P_T} \right]
\label{cphi-qht} \\
&& \sin \varphi_q^h = \frac{P_T}{\pp}\sin(\phi _h-\phi_\perp)\,.
\label{sphi-qht}
\eea
In Eq. (\ref{d++}) $D_{h/q}(z,p_\perp)$ is the unintegrated unpolarized
fragmentation function. Other common notations used in the literature are:
\be
\Delta^N  D_{h/q^\uparrow}(z,p_\perp) \equiv -2i\,D_{+-}^{h/q}(z,p_\perp) =
2 \,{\rm Im}D_{+-}^{h/q}(z,p_\perp) =
\frac{2p_{\perp}}{z M_h}H_{1}^{\perp}(z,p_{\perp})\,,
\label{D-prop}
\ee
referred to the Collins fragmentation function. $M_h$ is the mass of the produced
hadron.

\subsection{Elementary interaction}

The $\hat M_{\lambda_{\ell^\prime} \lambda_{q_f};\lambda_{\ell} \lambda_{q_i}}$ are the helicity amplitudes for the elementary process
$\ell \, q_i \to \ell^\prime q_f$, computed at LO in the $\gamma^* -p$~c.m.
frame, taking into account the quark intrinsic motion; the amplitudes are
normalized so that the unpolarized cross section, for a collinear collision,
is given by
\be
\frac{d\hat\sigma^{\ell q_i \to \ell^{\prime} q_f}}{d\hat t} =
\frac{1}{16\pi\hat s^2} \, \frac{1}{4} \sum_{\{\lambda\}}
|\hat M_{\lambda_{\ell ^\prime} \lambda_{q_f};\lambda_{\ell} \lambda_{q_i}}|^2\,,
\label{norm}
\ee
where $\hat t = -Q^2$ and $\hat s = \xb s$.

Helicity conservation for massless particles requires $\lambda_{\ell} =
\lambda_{\ell ^\prime}$, $\lambda_{q_i}=\lambda_{q_f} = \lambda_q$, which
implies that there are only two independent non-vanishing amplitudes,
explicitly computed in Appendix \ref{helicity-appendix}, with the result:
\bea
&&\hat M_1 \equiv \hat M_{++;++} = \hat M_{--;--}^* = e_q \, e^2
\left[\, \frac{1}{y} \, A_+ \, e^{+i\phi_\perp}\,
     -\, \frac{1-y}{y} \, A_- \, e^{-i\phi_\perp}
     -\,4\ \frac{\sqrt{1-y}}{y}\,\frac{\kt}{Q}\label{M++} \right] \\
&& \hat M_2 \equiv \hat M_{+-;+-} = \hat M_{-+;-+}^* = e_q \, e^2
\left[\, \frac{1-y}{y} \, A_+ \, e^{-i\phi_\perp}\,
     -\, \frac{1}{y} \, A_- \, e^{+i\phi_\perp}\,
     -\,4\, \frac{\sqrt{1-y}}{y}\,\frac{\kt}{Q}\label{M+-} \right] \,,
\eea
where $y=\frac{Q^2}{\xb s}$ and 
\be
A_{\pm} = \left( 1 \pm \sqrt{1+4\frac{\kt^2}{Q^2}} \right) \>.
\ee
These are exact LO results, holding at all orders in the $\kt/Q$ expansion.
By truncating this expansion at first order in $\kt/Q$, one obtains much
simpler expressions, which will be useful later,
\be
\hat M_1 = \hat M_{++;++} \simeq 2\,e_q e^2\left[
            \, \frac{1}{y}\,e^{+i\phi_\perp}\,
           -\,2\, \frac{\sqrt{1-y}}{y}\,\frac{\kt}{Q}\label{M++1} \right] \\
\ee
\be
\hat M_2 = \hat M_{+-;+-} \simeq 2\,e_q e^2\left[
            \, \frac{(1-y)}{y}\,e^{-i\phi_\perp}\,
           -\,2\, \frac{\sqrt{1-y}}{y}\,\frac{\kt}{Q}\label{M+-1} \right] \,\cdot
\ee
\vskip 18pt

We can now assemble the expression of the different factors - each
corresponding to a physical step - into Eqs. (\ref{gen-xs}) or
(\ref{app-xs}) to obtain the SIDIS cross section in terms of the TMDs.
This can be done in several ways. The most direct one is that of performing
the helicity sums in Eq.~(\ref{gen-xs}) taking into account
Eqs.~(\ref{rho-q}), (\ref{d++})--(\ref{d-+}),
(\ref{D-prop}), (\ref{M++}) and (\ref{M+-}). It yields:
\bea
\frac{d\sigma^{\ell(S_\ell) + p(S) \to \ell^\prime h X}}
{d\xb dQ^2 dz_h \, d^2 \bfP_T d\phi_S} &=& \frac {1}{2\pi}
\sum_{q}  \frac {1}{16 \, \pi \, (\xb s)^2}
\int d^2\bkt \, 
\frac {z}{z_h} \, J 
\nonumber \\
&\times& \!\! \frac{1}{2}\,\biggl\{ \hat{f}_{q/p,S}(x,\bkt )\,
\left( |\hat{M}_1|^2+|\hat{M}_2|^2 \right) \, D_{h/q}(z,p_{\perp})
\nonumber \\
&& + \, P_z^{\ell} \, P_z^{q} \, \hat{f}_{q/p,S}(x,\bkt) \,
\left( |\hat{M}_1|^2-| \hat{M}_2|^2\right) \, D_{h/q}(z,p_{\perp})
\label{kern2} \\
&& + \Bigl[ P_y^q \, \hat{f}_{q/p,S}(x,\bkt)
\left( {\rm Re} (\hat{M}_1\hat{M}^*_2) \cos\varphi_{q}^{h} -
{\rm Im} (\hat{M}_1\hat{M}^*_2) \sin\varphi_{q}^{h}\right) \nonumber\\
&& - \, P_x^q \, \hat{f}_{q/p,S}(x,\bkt)
\left( {\rm Im} (\hat{M}_1\hat{M}^*_2) \cos\varphi_{q}^{h} +
{\rm Re} (\hat{M}_1\hat{M}^*_2) \sin\varphi_{q}^{h}\right) \Bigr]
\Delta^N D_{h/q^{\uparrow}}(z,p_{\perp})
\biggr\} \>, \nonumber
\eea
which expresses the cross section in terms of the lepton and the quark
polarization vectors, the helicity amplitudes of the elementary interaction
and either the unpolarized or the Collins fragmentation functions.
The intrinsic transverse momentum of the produced hadron, $\bfp _{\perp}$,
is related to  $\bfk _{\perp}$ and the other kinematical variables as shown in
Eq.~(28) of Ref.~\cite{Anselmino:2005nn}. The exact expressions of
$\cos\varphi_{q}^{h}$ and $\sin\varphi_{q}^{h}$ can be obtained from
Eqs.~(C3) and (C4).

We now continue our computation, in this Section, at
${\cal O}\left( k_\perp /Q \right)$. From Eqs.~(\ref{M++1}), (\ref{M+-1}),
(\ref{cphi-qht}) and (\ref{sphi-qht}), we have:
\be
|\hat{M}_1|^2 + |\hat{M}_2|^2  =
\frac{4 e_q^2 e^4}{y^2}
\left[1+(1-y)^2 - 4(2-y)\sqrt{1-y}\,\frac{\kt}{Q}\cos\phi_\perp\right]
\label{M1+M2}
\ee
\vskip -12pt
\be
|\hat{M}_1|^2 - |\hat{M}_2|^2  =
\frac{4 e_q^2 e^4}{y^2}
\left[1-(1-y)^2 - 4y\sqrt{1-y}\,\frac{\kt}{Q}\cos\phi_\perp\right]
\label{M1-M2}
\ee
\vskip -12pt
\bea
{\rm Im} (\hat{M}_1\hat{M}^*_2) \, \cos\varphi_{q}^{h} +
{\rm Re} (\hat{M}_1\hat{M}^*_2) \, \sin\varphi_{q}^{h} &=&
\frac{P_T}{p_\perp} \frac{4 e_q^2 e^4}{y^2}
\left\{ (1-y) \left[ \sin(\phi_h +\phi_\perp) -
z_h \frac{\kt}{P_T} \sin2\phi_\perp\right]\right. \nonumber  \\
&-& \left.
2\sqrt{1-y}(2-y)\frac{\kt}{Q} \left[ \sin \phi_h -
z_h \frac{\kt}{P_T}\sin\phi_\perp \right] \right\}\label{imM1M2}
\eea
\vskip -12pt
\bea
{\rm Re} (\hat{M}_1\hat{M}^*_2) \, \cos\varphi_{q}^{h} -
{\rm Im} (\hat{M}_1\hat{M}^*_2) \, \sin\varphi_{q}^{h}
&=&
\frac{P_T}{p_\perp} \frac{4 e_q^2 e^4}{y^2}
\left\{ (1-y) \left[ \cos(\phi_h +\phi_\perp) -
z_h \frac{\kt}{P_T} \cos2\phi_\perp\right]\right. \nonumber  \\
&-& \left. 2\sqrt{1-y}(2-y)\frac{\kt}{Q} \left[
\cos\phi_h -  z_h \frac{\kt}{P_T}\cos\phi_\perp \right] \right\}
\cdot \label{reM1M2}
\eea

Inserting these results, together with Eqs.~(\ref{fqs})--(\ref{fyqs}), into
Eq.~(\ref{kern2}), gives, at order $k_\perp/Q$, the following expression for
the SIDIS cross section in the TMD factorization scheme:
\bea
&&
\frac{d\sigma^{\ell(S_\ell) + p(S) \to \ell^\prime h X}}
{d\xb \, dQ^2 \, dz_h \, d^2 \bfP_T \, d\phi_S} = \frac {1}{2\pi}
\sum_{q} \> \frac {1}{16 \, \pi \, (\xb s)^2} \int d^2\bkt \, d^2\bfp_\perp
\> \delta ^{(2)}(\bfP_T - z_h\bfk_\perp -\bfp_\perp)
\, \frac{4 e_q^2 e^4}{y^2}
\nonumber \\
&&
\biggl\{\,\frac{1}{2} f_{q/p} \left[ 1+(1-y)^2 \right] D_{h/q}
-\frac{1}{2}\Delta f^{q}_{s_y/p} \, \frac{P_T}{p_\perp} (1-y)
\left[ \cos(\phi_h + \phi_\perp) - z_h \, \frac{\kt}{P_T} \,
\cos2\phi_\perp \right] \Delta^N  D_{h/q^\uparrow} \nonumber  \\
&&
-2(2-y)\sqrt{1-y} \, \frac{\kt}{Q} \left[
f_{q/p}\cos\phi_\perp D_{h/q}
- \frac{1}{2}\Delta f^{q}_{s_y/p} \, \frac{P_T}{p_\perp}\left( \cos \phi_h
- z_h \, \frac{\kt}{P_T} \, \cos\phi_\perp \right)\Delta^N  D_{h/q^\uparrow}
\right] \nonumber \\
&&
+\frac{1}{2} S_L \left[\frac{P_T}{p_\perp}(1-y) \, \Delta f^{q}_{s_x/S_L}
\left( \sin(\phi_h +\phi_\perp) - z_h \, \frac{\kt}{P_T} \, \sin2\phi_\perp
\right) \Delta^N  D_{h/q^\uparrow} \nonumber \right. \\
&& \left.\hspace{0.8cm}
-2(2-y)\sqrt{1-y} \, \frac{\kt}{Q} \, \frac{P_T}{p_\perp} \,
\Delta f^{q}_{s_x/S_L} \left( \sin\phi_h - z_h \, \frac{\kt}{P_T} \,
\sin\phi_\perp \right) \Delta^N  D_{h/q^\uparrow} \nonumber \right. \\
&& \left.\hspace{0.8cm}
+P^\ell_z \left( \left[ 1-(1-y)^2 \right] \Delta f^{q}_{s_z/S_L} D_{h/q} -
4y\sqrt{1-y} \, \frac{\kt}{Q} \, \Delta f^{q}_{s_z/S_L} \cos\phi_\perp
D_{h/q} \right) \right]\nonumber \\
&&
+\frac{1}{2} S_T \left[
\frac{1}{2}\left[ 1+(1-y)^2 \right]\Delta f_{q/S_T}
\sin (\phi_\perp-\phi_S) D_{h/q} \nonumber \right. \\
&& \left.\hspace{0.8cm}
+ P^\ell_z\left[ 1-(1-y)^2 \right]\Delta f^{q}_{s_z/S_T}
\cos(\phi_\perp-\phi_S)D_{h/q}\nonumber \right. \\
&& \left.\hspace{0.8cm}
- P^\ell_z \, 2y\sqrt{1-y} \, \frac{\kt}{Q}\, \Delta f^{q}_{s_z/S_T}
\Big( \cos\phi_S + \cos(2\phi_\perp-\phi_S)\Big) D_{h/q}
\nonumber \right. \\
&& \left.\hspace{0.8cm}
+\frac{P_T}{2p_\perp}\,(1-y) \,
(\Delta f^{q}_{s_x/S_T} + \Delta ^- f^{q}_{s_y/S_T})
\Big( \sin(\phi_h+\phi_S)  - z_h \, \frac{\kt}{P_T} \,
\sin(\phi_\perp+\phi_S) \Big)  \Delta^N  D_{h/q^\uparrow}
\nonumber \right. \\
&& \left.\hspace{0.8cm}
+\frac{P_T}{2p_\perp}\,(1-y) \,
(\Delta f^{q}_{s_x/S_T} - \Delta ^- f^{q}_{s_y/S_T})
\Big( \sin(\phi_h + 2\phi_\perp - \phi_S)  - z_h \frac{\kt}{P_T}
\sin(3\phi_\perp - \phi_S) \Big)
\Delta^N  D_{h/q^\uparrow}
\nonumber \right. \\
&& \left.\hspace{0.8cm}
-\frac{P_T}{p_\perp}\,(2-y)\sqrt{1-y}\,\frac{\kt}{Q}
(\Delta f^{q}_{s_x/S_T} + \Delta ^- f^{q}_{s_y/S_T})
\Big( \sin(\phi_h-\phi_\perp + \phi_S)  - z_h \, \frac{\kt}{P_T} \,
\sin\phi_S \Big)  \Delta^N  D_{h/q^\uparrow}
\nonumber \right. \\
&& \left.\hspace{0.8cm}
-\frac{P_T}{p_\perp} \, (2-y)\sqrt{1-y}\,\frac{\kt}{Q}
(\Delta f^{q}_{s_x/S_T} - \Delta ^- f^{q}_{s_y/S_T})
\Big( \sin(\phi_h+\phi_\perp-\phi_S)  - z_h \, \frac{\kt}{P_T} \,
\sin(2\phi_\perp - \phi_S)\Big)  \Delta^N  D_{h/q^\uparrow}
\nonumber \right. \\
&& \left.\hspace{0.8cm}
+(2-y)\sqrt{1-y}\,\frac{\kt}{Q} \, \Delta f_{q/S_T} \Big(\sin \phi_S -
\sin (2\phi_\perp - \phi_S)\Big) D_{h/q}
\right] \biggr\}\,.
\label{xs-full-expl}
\eea

The first three terms of Eq.~(\ref{xs-full-expl}) correspond to the contribution
of the unpolarized proton to the SIDIS cross section; they contain either the
unpolarized or the Boer-Mulders distribution functions. The following three
terms correspond to the longitudinally-polarized proton contributions; they
depend either on the helicity distribution $\Delta f^{q}_{s_z/S_L} [= \Delta q = g_1]$
or on the $\Delta f^{q}_{s_x/S_L} [= (\kt/M) \, h^\perp_{1L}]$ transverse
momentum dependent distribution. Finally, the last eight terms correspond to
the transversely-polarized proton contributions; they may originate from the
Sivers function, from $\Delta f^{q}_{s_z/S_T} [= (\kt/M) \, g^\perp_{1T}]$, and
from the transversity distribution functions, related to the combinations
$(\Delta f^q_{s_x/S_T} \pm \Delta^- f^q_{s_y/S_T})$
as shown in Eqs.~(\ref{F+-+-}) and (\ref{F-++-}).
The partonic distributions couple either to the unpolarized or to the Collins
fragmentation functions, depending on whether they are, respectively, chiral
even or odd.

Notice that we have intentionally grouped all terms according to their phases,
so that this expression can be easily compared with the analogous formulae
of Ref.~\cite{Bacchetta:2006tn}, which have the same structure.
To make the comparison fully explicit, apart from converting
our notation to the Amsterdam group notation, we need to extract from the
integration over the intrinsic transverse momentum $\bkt$ the dependence on the
azimuthal angles $\phi_h$ and $\phi_S$. On the basis of a simple tensorial analysis,
which is described in detail in Appendices~\ref{t-An} and \ref{Int-p}, we can
recover Eqs.~(4.2)-(4.19) of Ref.~\cite{Bacchetta:2006tn}, without formulating
any particular assumption on the $x$ ($z$) and $\kt$ ($p_\perp$) dependence of
the distribution (fragmentation) functions.

In analogy with the Amsterdam notation, Ref.~\cite{Bacchetta:2006tn}, we define the
convolution on transverse momenta in the following way
\be
\mathcal{C}[w \, f\, D] =  \sum_{q} e_q^2 \int d^2\bkt \, d^2\bfp _\perp \,
\delta^{(2)}(\bfP_T - z_h \bkt - \bfp _\perp) \, w(\bkt, \bfP_T) \, f(\xb,\kt) \,
D\left(z_h,p_\perp \right)\; .
\ee
Notice that this definition differs from Eq.~(41) of Ref.~\cite{Bacchetta:2006tn}
by a factor $\xb$ and for the definition of the parton momenta, see
Eqs.~(\ref{a-noi1})--(\ref{a-noi3}).

The convolutions on intrinsic transverse momenta in the single terms of
Eq.~(\ref{xs-full-expl}) can in fact be written as:
\bea
F_{UU} &=& \sum_{q} e_q^2 \, \int d^2\bkt \, f_{q/p} \, D_{h/q} =
\mathcal{C}[f_1 \, D_1]
\label{FUU} \\
\cos2\phi_h\,F_{UU}^{\cos2\phi_h} &=& -\sum_{q} e_q^2 \, \int d^2\bkt \,
\Delta f^{q}_{s_y/p} \frac{P_T}{2\,p_\perp}
\left[ \cos(\phi_h + \phi_\perp) - z_h \frac{\kt}{P_T} \cos2\phi_\perp \right]
\Delta^N  D_{h/q^\uparrow} \nonumber \\
&=&\cos2\phi_h \; \mathcal{C}\left[
\frac{ (\bfP _T \cdot \bkt) - 2 z_h(\hat{\bfP}_T \cdot \bkt)^2 +
z_h \kt^2}{z_h M_h M} h_1^\perp \, H_1^\perp\right]
\label{FUUcos2phi} \\
\cos\phi_h\,F_{UU}^{\cos\phi_h} &=& -2\sum_{q} e_q^2 \, \int d^2\bkt
\,\frac{\kt}{Q} \Big\{ \cos\phi_\perp f_{q/p}  D_{h/q}
\Big.\nonumber \\
&& \hspace{3.2cm} \Big. - \frac{P_T}{2\,p_\perp} \left[ \cos\phi_h -
z_h \frac{\kt}{P_T} \cos\phi_\perp  \right]\Delta f^{q}_{s_y/p}
\Delta^N  D_{h/q^\uparrow}\Big\}\nonumber \\
&=&\cos\phi_h \left(-\frac{2}{Q}\right)  \; \mathcal{C}\left[
(\hat{\bfP}_T \cdot \bkt) f_1\,D_1 + \frac{
\kt^2\,\Big(P_T-z_h \,\hat{\bfP}_T \cdot \bkt\Big)}{z_h M_h M}
 \, h_1^\perp \, H_1^\perp \right]
\label{FUUcosphi}\\
\sin 2\phi_h\,F_{UL}^{\sin 2\phi_h} &=&\sum_{q} e_q^2 \, \int d^2\bkt
\,\frac{P_T}{2 p_\perp} \Delta f^{q}_{s_x/S_L}
\left( \sin(\phi_h +\phi_\perp) - z_h \frac{\kt}{P_T} \sin2\phi_\perp \right)
\Delta^N  D_{h/q^\uparrow} \nonumber \\
&=& \sin 2\phi_h  \; \mathcal{C}\left[
\frac{(\bfP _T \cdot \bkt) - 2 z_h (\hat{\bfP}_T \cdot \bkt)^2 + z_h \kt^2}{z_h\,M_h\,M}
 \, h_{1L}^\perp \, H_1^\perp \right]
\label{FULsin2phi}\\
\sin \phi_h\,F_{UL}^{\sin \phi_h} &=& -2\sum_{q} e_q^2 \, \int d^2\bkt
\, \frac{\kt}{Q}\frac{P_T}{2 p_\perp} \Delta f^{q}_{s_x/S_L}
\left( \sin\phi_h - z_h \frac{\kt}{P_T}\sin\phi_\perp \right)
\Delta^N  D_{h/q^\uparrow} \nonumber \\
&=& \sin\phi_h \left(-\frac{2}{Q}\right)  \; \mathcal{C}\left[
\frac{\kt^2 \Big(P_T -z_h(\hat{\bfP}_T \cdot \bkt)\Big)}{z_h\,M_h\,M}
\, h_{1L}^\perp \, H_1^\perp \right]
\label{FULsinphi}\\
\sin \phi_h\,F_{LU}^{\sin \phi_h} &=& 0 \hspace{.4cm}
{\rm at \;\;leading\;\; twist}
\label{FLUsinphi}\\
F_{LL} &=&\sum_{q} e_q^2 \, \int d^2\bkt \, \Delta f^{q}_{s_z/S_L} \, D_{h/q}
= \mathcal{C}[g_{1L} \, D_1]\label{FLL} \\
\cos \phi_h\,F_{LL}^{\cos \phi_h} &=&-2\sum_{q} e_q^2 \, \int d^2\bkt
\,\frac{\kt}{Q} \Delta f^{q}_{s_z/S_L} \cos\phi_\perp  D_{h/q} \nonumber \\
&=&  \cos \phi_h \left(-\frac{2}{Q}\right)     \; \mathcal{C}\left[
(\hat{\bfP}_T \cdot \bkt) g_{1L}  D_{1}\right]
\label{FLLcosphi}\\
\sin(\phi_h-\phi_S)\,F_{UT}^{\sin(\phi_h-\phi_S)} &=&
\frac{1}{2}\sum_{q} e_q^2 \, \int d^2\bkt \, \Delta f_{q/S_T}
\sin (\phi_\perp-\phi_S) D_{h/q} \nonumber \\
&=& \sin(\phi_h-\phi_S)      \; \mathcal{C}\left[
\frac{-(\hat{\bfP}_T \cdot \bkt )}{M}  \, f_{1T}^\perp \, D_1\right]
\label{FUTsin(phi-phiS)}\\
\cos(\phi_h-\phi_S)\,F_{LT}^{\cos(\phi_h-\phi_S)} &=& \sum_{q} e_q^2 \,
\int d^2\bkt \, \Delta f^q_{s_z/S_T} \cos (\phi_\perp-\phi_S) D_{h/q}
\nonumber \\
&=& \cos(\phi_h-\phi_S)     \; \mathcal{C}\left[
\frac{(\hat{\bfP}_T \cdot \bkt )}{M}  \, g_{1T}^\perp \, D_1\right]
\label{FLTcos(phi-phiS)} \\
\cos \phi_S\,F_{LT}^{\cos \phi_S} &=&
-\sum_{q} e_q^2 \, \int d^2\bkt  \frac{\kt}{Q} \Delta f^{q}_{s_z/S_T}
\cos \phi_S \,D_{h/q} \nonumber \\
&=& \cos \phi_S\,\left(-\frac{1}{Q}\right) \; \mathcal{C}\left[
\frac{\kt ^2}{M}  \, g_{1T}^\perp \, D_1\right]
\label{FUTcosphiS}\\
\cos(2\phi_h-\phi_S)\,F_{LT}^{\cos(2\phi_h-\phi_S)}&=&
-\sum_{q} e_q^2 \, \int d^2\bkt  \frac{\kt}{Q} \Delta f^{q}_{s_z/S_T}
\cos (2\phi_\perp-\phi_S) \,D_{h/q} \nonumber \\
&=&\cos(2\phi_h- \phi_S)\;\frac{1}{Q} \; \mathcal{C}\left[
\frac{ \Big(\kt^2 -
2(\hat{\bfP}_T \cdot \bkt)^2\Big)}{M} \, g_{1T}^\perp \, D_1 \right]
\label{FUTcos(2phih-phiS)} \\
\sin(\phi_h + \phi_S) F_{UT}^{\sin(\phi_h+\phi_S)} &=&
\sum_{q} e_q^2 \, \int d^2\bkt
\frac{P_T}{2p_\perp} \nonumber \\
&\times& (\Delta f^{q}_{s_x/S_T} + \Delta ^- f^{q}_{s_y/S_T})
\Big( \sin(\phi_h + \phi_S)  - z_h \frac{\kt}{P_T} \sin(\phi_\perp+\phi_S) \Big)
\Delta^N  D_{h/q^\uparrow}  \nonumber \\
&=&\sin(\phi_h + \phi_S)   \; \mathcal{C}\left[
\frac{ \left(P_T - z_h \kt (\hat{\bfP}_T \cdot \hat{\bfk}_\perp)\right)}{z_h\,M_h} \,
h_{1} \, H_1^\perp \right]
\label{FUTsin(phi+phiS)}\\
\sin(3\phi_h-\phi_S) F_{UT}^{\sin(3\phi_h-\phi_S)} &=&
\nonumber \\
&&\hspace{-3.7cm}=\sum_{q} e_q^2 \, \int d^2\bkt  \frac{P_T}{2p_\perp}
(\Delta f^{q}_{s_x/S_T} - \Delta ^- f^{q}_{s_y/S_T})
\Big( \sin(\phi_h + 2\phi_\perp - \phi_S) - z_h \frac{\kt}{P_T}
\sin(3\phi_\perp - \phi_S) \Big)
\Delta^N  D_{h/q^\uparrow}  \nonumber \\
&&\hspace{-3.7cm} =\sin(3\phi_h-\phi_S) \, \mathcal{C} \left[
\frac{{\kt^2}\Big\{-P_T + 2\,P_T\,(\hat{\bfP}_T \cdot \hat{\bfk}_\perp)^2 -
z_h {\kt}\left[4(\hat{\bfP}_T \cdot \hat{\bfk}_\perp)^3 +
3(\hat{\bfP}_T \cdot \hat{\bfk}_\perp) \right]\Big\}}{2 z_h\,M_h\,M^2 }
{h_{1T}^\perp} \, {H_1^\perp}  \right]
\label{FUTsin(3phi-phiS)}\\
\sin\phi_S F_{UT}^{\sin\phi_S} &=&
-\sum_{q} e_q^2 \, \int d^2\bkt \frac{P_T}{\pp}\,\frac{\kt}{Q}\nonumber \\
&& \times
(\Delta f^{q}_{s_x/S_T} + \Delta ^- f^{q}_{s_y/S_T})
\Big( \sin(\phi_h-\phi_\perp + \phi_S)  - z_h \frac{\kt}{P_T} \sin\phi_S \Big)
\Delta^N  D_{h/q^\uparrow}  \nonumber \\
&+& \frac{1}{2} \sum_{q} e_q^2 \, \int d^2\bkt  \frac{\kt}{Q}
\Delta f_{q/S_T} \sin \phi_S \,D_{h/q}  \nonumber \\
&=& \sin\phi_S \,\left(-\frac{2}{Q}\right)  \; \mathcal{C}\left[
\frac{\left(\bfP_T \cdot \bfk _\perp - z_h\kt^2\right)}{z_h\,M_h}  h_{1} \,
{H_1^\perp} \;
+ \; \frac{\kt ^2}{2M} f_{1T}^\perp \, D_1\right]
\label{FUTsinphiS}\\
\sin(2\phi_h-\phi_S) F_{UT}^{\sin(2\phi_h-\phi_S)} &=&
-\sum_{q} e_q^2 \, \int d^2\bkt \frac{P_T}{p_\perp}\,\frac{\kt}{Q}\nonumber \\
& & \times
(\Delta f^{q}_{s_x/S_T} - \Delta ^- f^{q}_{s_y/S_T})
\Big( \sin(\phi_h + \phi_\perp - \phi_S)  - z_h \frac{\kt}{P_T}
\sin(2\phi_\perp - \phi_S)\Big)
\Delta^N  D_{h/q^\uparrow}\nonumber \\
& - & \frac{1}{2}\sum_{q} e_q^2 \, \int d^2\bkt  \frac{\kt}{Q}
\Delta f_{q/S_T} \sin (2\phi_\perp - \phi_S) \,D_{h/q} \nonumber \\
&=& \sin(2\phi_h-\phi_S)\,\left(-\frac{1}{Q}\right)\,  \; \mathcal{C}\Big[
  {\frac{ {\kt^2} \left( (\bfP_T \cdot \bfk_\perp) +
z_h \kt^2 \left(1 - 2(\hat{\bfP}_T \cdot \hat{\bfk}_\perp)^2 \right)\right) }{z_h\,M_h\,M^2}
h_{1T}^\perp \, {H_1^\perp}} \nonumber \\
&& \hspace{5.0cm}  - \frac{\left( 2 (\hat{\bfP}_T \cdot \bkt)^2 -
{\kt^2} \right)}{M}\,{f_{1T}^\perp} \, D_1 \Big]\,.
\label{FUTsin(phi-2phiS)}
\eea
These ``$F_{S_\ell S}$ structure functions" are the same as those defined in
Ref.~\cite{Bacchetta:2006tn}, apart from an overall factor $\xb$ which
appears in the latter. In the comparison one should consider only leading
twist TMDs and remember the different notations of Ref.~\cite{Bacchetta:2006tn},
Eqs. (\ref{a-noi1})--(\ref{a-noi3}). Using the above $F$'s
in Eq.~(\ref{xs-full-expl}) one obtains the full expression of the SIDIS
polarized cross section, valid with leading twist TMDs and at kinematical order
$\kt/Q$:
\bea
\frac{d\sigma^{\ell(S_\ell) + p(S) \to \ell^\prime h X}}
{d\xb \, dQ^2 \, dz_h \, d^2 \bfP_T \, d\phi_S}
&=& \frac {2 \, \alpha^2}{Q^4} \nonumber \\
&\times&  \bigg\{
\frac{1+(1-y)^2}{2} F_{UU} + (2-y)\sqrt{1-y} \, \cos\phi_h \,
F_{UU}^{\cos\phi_h} + (1-y) \, \cos2\phi_h \, F_{UU}^{\cos2\phi_h}\Big.
\nonumber \\
&& + \, \Big. S_L \,\Big[ (1-y) \, \sin 2\phi_h \, F_{UL}^{\sin2\phi_h}  +
(2-y)\sqrt{1-y} \, \sin\phi_h \, F_{UL}^{\sin\phi_h} \Big]\Big.
\nonumber \\
&& + \, \Big. S_L \, P_z^\ell \, \Big[ \frac{1-(1-y)^2}{2} \, F_{LL} +
y\sqrt{1-y} \, \cos\phi_h \, F_{LL}^{\cos\phi_h} \Big]
\nonumber\Big. \\
&& + \, \Big. S_T \, \Big[ \frac{1+(1-y)^2}{2} \, \sin(\phi_h-\phi_S) \,
F_{UT}^{\sin(\phi_h-\phi_S)} \Big.\Big.\nonumber \\
&& \Big.\hspace{0.8cm} +(1-y) \, \Big( \sin(\phi_h + \phi_S) \,
F_{UT}^{\sin(\phi_h + \phi_S)} + \sin(3\phi_h-\phi_S) \,
F_{UT}^{\sin(3\phi_h-\phi_S)} \Big) \Big. \Big.\nonumber \\
&& \hspace{0.8cm} \, + \Big. \Big. \, (2-y) \, \sqrt{1-y} \,
\Big( \sin\phi_S \, F_{UT}^{\sin\phi_S} +
\sin(2\phi_h-\phi_S) \, F_{UT}^{\sin(2\phi_h-\phi_S)} \Big) \Big] \Big.
\nonumber \\
&& + \, \Big. S_T \, P_z^\ell \, \Big[  \frac{1-(1-y)^2}{2} \, \cos(\phi_h-\phi_S) \, F_{LT}^{\cos(\phi_h-\phi_S)} \Big.\Big. \nonumber \\
&& \Big. \hspace{1.2cm} +  y\sqrt{1-y} \, \Big(\cos\phi_S \, F_{LT}^{\cos\phi_S} +
\cos(2\phi_h-\phi_S) \, F_{LT}^{\cos(2\phi_h-\phi_S)} \Big) \Big] \bigg\} \>.
\label{UT}
\eea
This expression agrees with Eq.~(2.7) of Ref.~\cite{Bacchetta:2006tn}, bearing
in mind  Eqs.~(2.8--2.13) and that, at leading twist,
$F_{UU,L} = F_{LU}^{\sin\phi_h} = 0$.

In obtaining the general cross section structure of Eq. (\ref{UT}) we started
from the TMD factorization, Eq. (\ref{gen-xs}); then we have simply exploited
the properties of the helicity amplitudes, which essentially originate from the
phase dependence of the Dirac spinors and their non collinear kinematics. Each
step of the factorization scheme contributes some phases, including the
elementary interactions.

Some of the final azimuthal dependences have a clear and direct physical
interpretation. For example, the phase of $F_{UT}^{\sin(\phi_h-\phi_S)}$,
Eq. (\ref{FUTsin(phi-phiS)}), originates from the phase dependence of the
$\Delta \hf_{{q}/S_T}(x, \bkt)$ distribution, Eq.~(\ref{DD1}). This is the
Sivers effect \cite{Sivers:1989cc,Sivers:1990fh},
which relates the number of unpolarized quarks with intrinsic
momentum $\bkt$ to the spin of the proton; such an effect, due to parity
invariance, can only be of the form $\bfS \cdot (\hat{\bfp} \times
\hat{\bfk}_\perp) = S_T \, \sin(\phi_\perp -\phi_S)$. Similarly, the phase
in the first term of $F_{UU}^{\cos\phi_h}$, Eq. (\ref{FUUcosphi}), being
associated with unpolarized distribution and fragmentation functions, can only
come from the $\bkt$ dependence of the elementary interaction, the so-called
Cahn effect \cite{Anselmino:2005nn}.

\section{\label{asy}Single and double spin asymmetries in SIDIS}

From the expression of the SIDIS polarized cross section we can now compute
all spin asymmetries which have been, or can be, measured. We can restart from Eq.~(\ref{kern2}), inserting into it the expressions of the polarized quark
distributions, as given in Eqs.~(\ref{fqs})--(\ref{DD7}):
\bea
&&\frac{d\sigma^{\ell(S_\ell) + p(S) \to \ell^\prime h X}}
{d\xb \, dQ^2 \, dz_h \, d^2 \bfP_T \, d\phi_S} = \frac {1}{2\pi}
\sum_{q}  \frac {1}{16 \, \pi \, (\xb s)^2}
\int d^2\bkt \, 
\frac {z}{z_h} \, J 
\nonumber \\
&\times& \!\! \frac{1}{2}\,\Biggl\{ \left(
f_{q/p}(x, \kt) + \frac{1}{2} S_T \,\Delta \hat f_{q/S_T}(x, \bkt) \right)
\left(|\hat{M}_1|^2+|\hat{M}_2|^2 \right) \, D_{h/q}(z,p_{\perp})
\nonumber\\
&+& P_z^{\ell} \left(
S_L \,\Delta \hat f^{q}_{s_z/S_L}(x, \bkt) + S_T \,
\Delta \hat f^{q}_{s_z/S_T}(x, \bkt) \right)
\left(|\hat{M}_1|^2-| \hat{M}_2|^2\right) \, D_{h/q}(z,p_{\perp})
\label{gen-xs-asy} \\
&-& \biggl[ \left(
 \Delta f^{q}_{s_y/p}(x, \kt) - S_T \, \Delta ^- \hat f^{q}_{s_y/S_T}(x, \bkt)
\right) \left( {\rm Re} (\hat{M}_1\hat{M}^*_2) \cos\varphi_{q}^{h} -
{\rm Im} (\hat{M}_1\hat{M}^*_2) \sin\varphi_{q}^{h} \right)
\nonumber\\
&& + \left(
S_L \,\Delta f^{q}_{s_x/S_L}(x, \kt) + S_T \,
\Delta \hat f^{q}_{s_x/S_T}(x, \bkt) \right)
\left( {\rm Im} (\hat{M}_1\hat{M}^*_2) \cos\varphi_{q}^{h} +
{\rm Re} (\hat{M}_1\hat{M}^*_2) \sin\varphi_{q}^{h}\right) \biggr]
\Delta^N D_{h/q^{\uparrow}}(z,p_{\perp})
\Biggr\} \, \cdot \nonumber
\eea
Notice that this expression, at leading twist, is exact at all orders in $\kt/Q$. We list here some
properties of the polarized distribution functions which are useful in computing
the asymmetries \cite{Anselmino:2005sh}:
\bea
&&\hat f_{q/S_T}(x,\bkt) + \hat f_{q/-S_T}(x,\bkt) = 2 f_{q/p}(x,\kt)
\nonumber \\
&&\hat f_{q/S_T}(x,\bkt) - \hat f_{q/-S_T}(x,\bkt) =
\Delta \hat f_{q/S_T}(x,\bkt) \nonumber \\
&& \Delta \hat f_{s_x/S_T}(x,\bkt) = -\Delta \hat f_{s_x/-S_T}(x,\bkt)
\nonumber \\
&& \Delta \hat f_{s_y/S_T}(x,\bkt) - \Delta \hat f_{s_y/-S_T}(x,\bkt) =
2\,\Delta^- \hat f^{q}_{s_y/S_T}(x,\bkt) \label{prop} \\
&& \Delta \hat f_{s_y/S_T}(x,\bkt) + \Delta \hat f_{s_y/-S_T}(x,\bkt) =
-2\,\Delta f^{q}_{s_y/p}(x,\kt) \nonumber \\
&& \Delta \hat f_{s_z/S_T}(x,\bkt) = -\Delta \hat f_{s_z/-S_T}(x,\bkt)
\nonumber \\
&& \Delta \hat f_{s_i/S_L}(x,\bkt) = \Delta \hat f_{s_i/-S_L}(x,\bkt)
\quad\quad (i=x,y,z)\,.
\nonumber
\eea

Let us now consider Eq.~(\ref{gen-xs-asy}) in several particular cases.
In the sequel, transverse and longitudinal always refer, both for the
protons and the leptons, to their (different) directions of motion in the
$\gamma^* - p$ c.m. frame. Longitudinal states coincide with helicity states.

\subsection{\label{A_UT} Nucleon transverse single spin asymmetry, $A_{UT}$}

Let us start with one of the most common SIDIS single spin asymmetries, $A_{S_\ell S}$, with
unpolarized leptons (U) and transversely polarized protons (T):
\be
A_{UT}\equiv
\frac{d^6\sigma^{\ell p^\uparrow \to \ell^\prime h X} -
      d^6\sigma^{\ell p^\downarrow \to \ell^\prime h X}}
     {d^6\sigma^{\ell p^\uparrow \to \ell^\prime h X} +
      d^6\sigma^{\ell p^\downarrow \to \ell^\prime h X}}=
\frac{d^6\sigma^{\ell + p(S_T) \to \ell^\prime h X} -
      d^6\sigma^{\ell + p(-S_T) \to \ell^\prime h X}}
     {d^6\sigma^{\ell + p(S_T) \to \ell^\prime h X} +
      d^6\sigma^{\ell + p(-S_T) \to \ell^\prime h X}} \cdot
\ee
For the numerator of $A_{UT}$ we have:
\bea
\!\!\!\!\!\!\!\!\!
\frac{d\sigma^{\ell p^\uparrow \to \ell^\prime h X} -
d\sigma^{\ell p^\downarrow \to \ell^\prime h X}}
{d\xb \, dQ^2 \, dz_h \, d^2 \bfP _T \, d\phi_S} &=& \frac {1}{2\pi}
\sum_{q}  \frac {1}{16 \, \pi \, (\xb s)^2}
\int d^2\bkt \, 
\frac {z}{z_h} \, J 
\label{num} \\
&\times& \biggl\{ \frac{1}{2}\Delta \hat f_{q/S_T}(x, \bkt) \,
(|{\hat M}_1|^2 +|{\hat M}_2|^2 ) \,
D_{h/q}(z, p_\perp) \nonumber \\
&+& \left[ \Delta^- \hat f^{q}_{s_y/S_T}(x, \bkt)
\left({\rm Re}({\hat M}_1{\hat M}_2^*) \, \cos\phi_{q}^h -
{\rm Im}({\hat M}_1{\hat M}_2^*) \, \sin\phi_{q}^h) \right) \right. \nonumber \\
&& - \left. \Delta \hat f^{q}_{s_x/S_T}(x, \bkt) \left(
{\rm Re}({\hat M}_1{\hat M}_2^*) \, \sin\phi_{q}^h +
{\rm Im}({\hat M}_1{\hat M}_2^*) \, \cos\phi_{q}^h) \right)\right]
\Delta^N  D_{h/q^\uparrow}(z, p_\perp) \biggr\}\,\cdot \nonumber
\eea
The first term in Eq.~(\ref{num}) corresponds to the Sivers effect,
whereas the second and the third terms correspond to the Collins effect,
coupled to the transversity distributions.

Similarly, for the denominator we find:
\bea
\!\!\!\!\!\frac{d\sigma^{\ell p^\uparrow \to \ell^\prime h X} +
d\sigma^{\ell p^\downarrow \to \ell^\prime h X}}
{d\xb \, dQ^2 \, dz_h \, d^2 \bfP _T \, d\phi_S} &=& \frac {1}{2\pi}
\sum_{q}  \frac {1}{16 \, \pi \, (\xb s)^2}
\int d^2\bkt \, 
\frac {z}{z_h} \, J 
\nonumber \\
&\times& \biggl\{
f_{q/p}(x, \kt) \, (|{\hat M}_1|^2 +|{\hat M}_2|^2 ) \, D_{h/q} (z, p_\perp)
\nonumber \\
&& -\,\Delta f^{q}_{s_y/p}(x, \kt) \left(
{\rm Re}({\hat M}_1{\hat M}_2^*) \, \cos\phi_{q}^h -
{\rm Im}({\hat M}_1{\hat M}_2^*) \, \sin\phi_{q}^h \right)
\Delta^N  D_{h/q^\uparrow}\biggr\}\,\cdot
\label{den}
\eea
Here, the first term corresponds to the usual unpolarized cross section (which
survives in the collinear limit) whereas the second term is an effect obtained
combining the Boer-Mulders distribution function, $\Delta f^{q}_{s_y/p}(x,\kt)$,
with the Collins fragmentation function, $\Delta^N  D_{h/q^\uparrow}(z,\pp)$.

If we insert the exact relations for ${\hat M}_1$ and ${\hat M}_2$ -- given
in Eqs.~(\ref{M++}) and (\ref{M+-}) -- and for $\cos\varphi^h_{q},\,\sin\varphi^h_{q}$
-- given in Eq.~(\ref{phi-qh}) -- into Eqs.~(\ref{num}) and (\ref{den}), we
obtain an exact expression for the $A_{UT}$ asymmetry. As already mentioned,
the numerator is given by two different contributions, the Sivers and the
Collins effect. Similarly, the denominator, which is simply twice the
unpolarized cross section for the $\ell \, p \to \ell^\prime h X$ process,
receive most contribution from the first term, proportional to the unpolarized distribution and fragmentation functions, with a further contribution from a
combination of the Boer-Mulders and Collins effects.

Much simpler, and often quite accurate, expressions can be obtained at
$\mathcal{O}(k_\perp/Q)$, neglecting higher order corrections. Using
Eqs.~(\ref{M1+M2})--(\ref{reM1M2}) and (\ref{DD1})--(\ref{DD7}) in
Eqs. (\ref{num}) and (\ref{den}), one has:
\bea
&&\frac{d\sigma^{\ell p^\uparrow \to \ell^\prime h X} -
d\sigma^{\ell p^\downarrow \to \ell^\prime h X}}
{d\xb \, dQ^2 \, dz_h \, d^2 \bfP_T \, d\phi_S} = \nonumber \\
&&\hspace{0.8cm}
\frac{2 \alpha^2}{Q^4}
\label{num1} \sum_{q} e_q^2 \int d^2\bkt
\left\{
\frac{1}{2}\Delta f_{q/S_T} \sin(\phi_\perp-\phi_S)
[ 1+(1-y)^2 ] \,D_{h/q} \right. \nonumber \\
&&\left.\hspace{0.8cm}
+\frac{P_T}{2p_\perp}\,(1-y) \,
(\Delta f^{q}_{s_x/S_T} + \Delta ^- f^{q}_{s_y/S_T})
\Big( \sin(\phi_h+\phi_S)  - z_h \, \frac{\kt}{P_T} \,
\sin(\phi_\perp+\phi_S) \Big)  \Delta^N  D_{h/q^\uparrow}
\nonumber \right. \\
&& \left.\hspace{0.8cm}
+\frac{P_T}{2p_\perp}\,(1-y) \,
(\Delta f^{q}_{s_x/S_T} - \Delta ^- f^{q}_{s_y/S_T})
\Big( \sin(\phi_h + 2\phi_\perp - \phi_S)  - z_h \frac{\kt}{P_T}
\sin(3\phi_\perp - \phi_S) \Big)
\Delta^N  D_{h/q^\uparrow}
\nonumber \right. \\
&& \left.\hspace{0.8cm}
-\frac{P_T}{p_\perp}\,(2-y)\sqrt{1-y}\,\frac{\kt}{Q}
(\Delta f^{q}_{s_x/S_T} + \Delta ^- f^{q}_{s_y/S_T})
\Big( \sin(\phi_h-\phi_\perp+\phi_S)  - z_h \, \frac{\kt}{P_T} \,
\sin\phi_S \Big)  \Delta^N  D_{h/q^\uparrow}
\nonumber \right. \\
&& \left.\hspace{0.8cm}
-\frac{P_T}{p_\perp} \, (2-y)\sqrt{1-y}\,\frac{\kt}{Q}
(\Delta f^{q}_{s_x/S_T} - \Delta ^- f^{q}_{s_y/S_T})
\Big( \sin(\phi_h+\phi_\perp-\phi_S)  - z_h \, \frac{\kt}{P_T} \,
\sin(2\phi_\perp - \phi_S)\Big)  \Delta^N  D_{h/q^\uparrow}
\nonumber \right. \\
&& \left.\hspace{0.8cm}
+(2-y)\sqrt{1-y}\,\frac{\kt}{Q} \, \Delta f_{q/S_T} \Big(\sin \phi_S -
\sin (2\phi_\perp-\phi_S)\Big) D_{h/q}
 \right\}\,.
\nonumber \\
&& \hspace{0.3cm} =
\frac{2 \alpha^2}{Q^4} \left\{
\left[ 1+(1-y)^2 \right]\sin(\phi_h-\phi_S)
F_{UT}^{\sin(\phi_h-\phi_S)} \label{num1-s} \right.\\
&& \left.\hspace*{1.6cm}
+ \,2(1-y)\left[ \sin(\phi_h+\phi_S) \, F_{UT}^{\sin(\phi_h+\phi_S)} +
\sin(3\phi_h-\phi_S) \, F_{UT}^{\sin(3\phi_h-\phi_S)} \right] \right. \nonumber \\
&& \left. \hspace*{1.6cm}
+ \, 2(2-y)\sqrt{1-y}\left[ \sin\phi_S \, F_{UT}^{\sin\phi_S} +
\sin(2\phi_h-\phi_S) \, F_{UT}^{\sin(2\phi_h-\phi_S)}\right]\right\} \nonumber
\eea
and
\bea
&&\frac{d\sigma^{\ell p^\uparrow \to \ell^\prime h X} +
d\sigma^{\ell p^\downarrow \to \ell^\prime h X}}
{d\xb \, dQ^2 \, dz_h \, d^2 \bfP_T \, d\phi_S} = \nonumber \\
&& \hspace*{0.2cm} \frac{2 \alpha^2}{Q^4}
\sum_{q} e_q^2 \int d^2\bkt \label{den1}
\biggl\{ f_{q/p} \left[ 1+(1-y)^2-4(2-y) \, \sqrt{1-y} \,
\frac{\kt}{Q} \,\cos\phi_\perp \right] \, D_{h/q} \nonumber \\
&& \hspace*{0.2cm} - \> \Delta f^{q}_{s_y/p} \left[
(1-y)\left(\cos(\phi_h+\phi_\perp)
-z_h \, \frac{\kt}{P_T} \, \cos(2\phi_\perp) \right) \nonumber \right.\\
&& \left.\hspace*{1.8cm}
-2(2-y) \, \sqrt{1-y} \, \frac{\kt}{Q}
\left(\cos\phi_h -z_h \, \frac{\kt}{P_T} \, \cos\phi_\perp \right)
\right]\frac{P_T}{\pp} \, \Delta^N  D_{h/q^\uparrow} \biggr\}
\nonumber \\
&=&
\frac{2 \alpha^2}{Q^4} \biggl\{ \left[ 1+(1-y)^2 \right] \, F_{UU}
+ 2(1-y)\cos 2\phi_h \, F_{UU}^{\cos 2\phi_h} + 2(2-y)\sqrt{1-y} \,
\cos\phi_h \, F_{UU}^{\cos\phi_h} \biggr\} \>, \label{den1-s}
\eea
where we have also exploited the definitions of the $F$ structure functions,
Eqs.~(\ref{FUU})--(\ref{FUTsin(phi-2phiS)}). These last expressions,
Eqs. (\ref{num1-s}) and (\ref{den1-s}), can also be obtained directly from
Eq.~(\ref{UT}). We recall that, at $\mathcal{O}(k_\perp/Q)$, one has $x = \xb$,
$z = z_h$, $\bfp_\perp = \bfP _T - z_h \bfk _\perp$ and $J=1$.

The first term in Eq.~(\ref{num1}) corresponds to the SIDIS Sivers asymmetry, 
which we analyzed in Refs.~\cite{Anselmino:2008sga,Anselmino:2005ea,Anselmino:2005nn,Anselmino:2010bs} 
for the extraction of the Sivers function, while the second term corresponds to the SIDIS Collins asymmetry, 
studied in Refs.~\cite{Anselmino:2007fs,Anselmino:2008jk} and used for the simultaneous extraction of the Collins and
transversity functions.

\subsection{\label{A_UL} Nucleon longitudinal single spin asymmetry, $A_{UL}$}

This asymmetry is defined for unpolarized leptons and a longitudinally
polarized proton target:
\be
A_{UL}\equiv
\frac{d^6\sigma^{\ell p^\rightarrow \to \ell^\prime h X} -
      d^6\sigma^{\ell p^\leftarrow \to \ell^\prime h X}}
     {d^6\sigma^{\ell p^\rightarrow \to \ell^\prime h X} +
      d^6\sigma^{\ell p^\leftarrow \to \ell^\prime h X}}=
\frac{d^6\sigma^{\ell + p(S_L) \to \ell^\prime h X} -
      d^6\sigma^{\ell + p(-S_L) \to \ell^\prime h X}}
     {d^6\sigma^{\ell + p(S_L) \to \ell^\prime h X} +
      d^6\sigma^{\ell + p(-S_L) \to \ell^\prime h X}}\,.
\ee
We give explicit results, for this and the next asymmetries, only valid at
$\mathcal{O}(k_\perp/Q)$. The denominator, as in the previous asymmetry, is
twice the unpolarized cross section and is given in Eq.~(\ref{den1}). For
the numerator we have:
\bea
\frac{d\sigma^{\ell + p(S_L) \to \ell^\prime h X} -
d\sigma^{\ell + p(-S_L) \to \ell^\prime h X}}
{d\xb \, dQ^2 \, dz_h \, d^2 \bfP_T \, d\phi_S} =
\frac{4 \, \alpha^2}{Q^4} \left\{
(1-y)\sin2\phi_h F_{UL} ^{\sin2\phi_h}
+ \sqrt{1-y}(2-y)\sin\phi_h F_{UL}^{\sin\phi_h}\right\} \>, \label{denUL-s}
\eea
as can be easily checked from Eq.~(\ref{UT}).

\subsection{\label{T-A_LL} Nucleon longitudinal double spin asymmetry, $A_{LL}$}

This asymmetry is defined by keeping fixed the longitudinal polarization of
the lepton, while flipping the direction of the proton target longitudinal
polarization:
\be
A_{LL}=
\frac{d^6\sigma^{\ell^\rightarrow p^\rightarrow \to \ell^\prime h X} -
      d^6\sigma^{\ell^\rightarrow p^\leftarrow \to \ell^\prime h X}}
     {d^6\sigma^{\ell^\rightarrow p^\rightarrow \to \ell^\prime h X} +
      d^6\sigma^{\ell^\rightarrow p^\leftarrow \to \ell^\prime h X}}=
\frac{d^6\sigma^{\ell (S_\ell) + p(S_L) \to \ell^\prime h X} -
      d^6\sigma^{\ell (S_\ell) + p(-S_L) \to \ell^\prime h X}}
     {d^6\sigma^{\ell (S_\ell) + p(S_L) \to \ell^\prime h X} +
      d^6\sigma^{\ell (S_\ell) + p(-S_L) \to \ell^\prime h X}}\,.
\ee
The denominator is the same as given in Eq.~(\ref{den1}), while for the numerator
we have
\bea
&&\hspace*{-0.3cm}\frac{d\sigma^{\ell (S_\ell) + p(S_L) \to \ell^\prime h X} -
d\sigma^{\ell (S_\ell) + p(-S_L) \to \ell^\prime h X}}
{d\xb \, dQ^2 \, dz_h \, d^2\bfP_T \, d\phi_S} = \nonumber \\
&&\hspace*{-0.3cm}
\frac{2 \alpha^2}{Q^4}  \left\{ \! [1-(1-y)^2] F_{LL} + 2y\sqrt{1-y}\,\cos\phi_hF_{LL}^{\cos\phi_h}
+2(1-y)\sin2\phi_h F_{UL} ^{\sin2\phi_h}
+ 2(2-y)\sqrt{1-y}\sin\phi_h F_{UL}^{\sin\phi_h}\right\}. \nonumber \\ \label{numLL}
\eea

\subsection{\label{l-A_LL} Lepton longitudinal double spin asymmetry,
$\tilde{A}_{LL}$}

This asymmetry is defined by keeping fixed the longitudinal polarization of the
proton target, while flipping the lepton longitudinal polarization:
\be
\tilde{A}_{LL}=
\frac{d^6\sigma^{\ell^\rightarrow p^\rightarrow \to \ell^\prime h X} -
      d^6\sigma^{\ell^\leftarrow p^\rightarrow \to \ell^\prime h X}}
     {d^6\sigma^{\ell^\rightarrow p^\rightarrow \to \ell^\prime h X} +
      d^6\sigma^{\ell^\leftarrow p^\rightarrow \to \ell^\prime h X}}=
\frac{d^6\sigma^{\ell (S_\ell) + p(S_L) \to \ell^\prime h X} -
      d^6\sigma^{\ell (-S_\ell) + p(S_L) \to \ell^\prime h X}}
     {d^6\sigma^{\ell (S_\ell) + p(S_L) \to \ell^\prime h X} +
      d^6\sigma^{\ell (-S_\ell) + p(S_L) \to \ell^\prime h X}}\,.
\ee
For the numerator we have
\bea
\frac{d\sigma^{\ell (S_\ell) + p(S_L) \to \ell^\prime h X} -
d\sigma^{\ell (-S_\ell) + p(S_L) \to \ell^\prime h X}}
{d\xb \, dQ^2 \, dz_h \, d^2\bfP_T \, d\phi_S} =
\frac{2 \alpha^2}{Q^4}  \left\{[1-(1-y)^2] F_{LL} + 2y\sqrt{1-y}\,\cos\phi_h F_{LL}^{\cos\phi_h}
\right\} \,.\nonumber \\ \label{l-numLL}
\eea
Notice that, in this case, the denominator differs from that given in
Eqs.~(\ref{den1}), as it acquires additional terms
(generated by $\Delta f^{q}_{s_x/S_L}$):
\bea
&&\frac{d\sigma^{\ell (S_\ell) + p(S_L) \to \ell^\prime h X} + d\sigma^{\ell (-S_\ell) + p(S_L) \to \ell^\prime h X}}
{d\xb \, dQ^2 \, dz_h \, d^2 \bfP_T \, d\phi_S} = \nonumber \\
&&\frac{2 \alpha^2}{Q^4} \biggl\{ \left[ 1+(1-y)^2 \right] \, F_{UU}
+ 2(1-y)[\cos 2\phi_h \, F_{UU}^{\cos 2\phi_h} +\sin 2\phi_h \, F_{UL}^{\sin 2\phi_h} ] \biggr. \nonumber \\
&& \hspace*{0.9cm} \biggl.
+\>  2(2-y)\sqrt{1-y} \,
[\cos\phi_h \, F_{UU}^{\cos\phi_h} +\sin \phi_h \, F_{UL}^{\sin \phi_h} ]
\biggr\} \>. 
\eea

\subsection{\label{t-A_LT} Nucleon longitudinal-transverse double spin asymmetry,
$A_{LT}$}

This asymmetry is defined by keeping fixed the longitudinal polarization of the lepton, while flipping the proton target transverse polarization:
\be
A_{LT}=
\frac{d^6\sigma^{\ell^\rightarrow p^\uparrow \to \ell^\prime h X} -
      d^6\sigma^{\ell^\rightarrow p^\downarrow \to \ell^\prime h X}}
     {d^6\sigma^{\ell^\rightarrow p^\uparrow \to \ell^\prime h X} +
      d^6\sigma^{\ell^\rightarrow p^\downarrow \to \ell^\prime h X}}=
\frac{d^6\sigma^{\ell (S_\ell) + p(S_T) \to \ell^\prime h X} -
      d^6\sigma^{\ell (S_\ell) + p(-S_T) \to \ell^\prime h X}}
     {d^6\sigma^{\ell (S_\ell) + p(S_T) \to \ell^\prime h X} +
      d^6\sigma^{\ell (S_\ell) + p(-S_T) \to \ell^\prime h X}}\,\cdot
\ee
The denominator is given in Eq.~(\ref{den1}), while for the numerator we have
\bea
&&\frac{d\sigma^{\ell (S_\ell) + p(S_T) \to \ell^\prime h X} -
d\sigma^{\ell (S_\ell) + p(-S_T) \to \ell^\prime h X}}
{d\xb \, dQ^2 \, dz_h \, d^2\bfP_T \, d\phi_S} = \nonumber \\
&&
\frac{2 \alpha^2}{Q^4} \left\{
\Big[1+(1-y)^2\Big]\sin(\phi_h-\phi_S) F_{UT}^{\sin(\phi_h-\phi_S)}
\right. \nonumber \\
&& \left. \hspace*{0.5cm}
+\Big[1-(1-y)^2\Big]\cos(\phi_h-\phi_S) F_{LT}^{\cos(\phi_h -\phi_S)}
\right. \nonumber \\
&& \left. \hspace*{0.5cm}
+ 2y\sqrt{1-y}\Big[\cos\phi_S F_{LT}^{\cos\phi_S} + \cos(2\phi_h-\phi_S)F_{LT}^{\cos(2\phi_h-\phi_S)}\Big]
\right.\nonumber \\
&& \left. \hspace*{0.5cm}
+ 2(1-y)\Big[\sin(\phi_h+\phi_S) F_{UT}^{\sin(\phi_h+\phi_S)} + \sin(3\phi_h-\phi_S)F_{UT}^{\sin(3\phi_h-\phi_S)}\Big]
\right.\nonumber \\
&& \left. \hspace*{0.5cm}
+ 2(2-y)\sqrt{1-y}\Big[\cos\phi_S F_{LT}^{\cos\phi_S} + \cos(\phi_h-\phi_S)F_{LT}^{\cos(\phi_h-\phi_S)}  \Big.
\right.\nonumber \\
&& \left. \Big. \hspace*{3.0cm}
+\sin\phi_S F_{UT}^{\sin\phi_S} + \sin(2\phi_h-\phi_S)F_{UT}^{\sin(2\phi_h-\phi_S)}\Big]\right\}\,.
\label{numLT}
\eea

\subsection{\label{l-A_LT} Lepton  longitudinal-transverse double spin asymmetry
$\tilde{A}_{LT}$}

This asymmetry is defined by flipping the direction of the longitudinal
polarization of the lepton, while keeping fixed the proton target transverse
polarization:
\be
\tilde{A}_{LT}=
\frac{d^6\sigma^{\ell^\rightarrow p^\uparrow \to \ell^\prime h X} -
      d^6\sigma^{\ell^\leftarrow p^\uparrow \to \ell^\prime h X}}
     {d^6\sigma^{\ell^\rightarrow p^\uparrow \to \ell^\prime h X} +
      d^6\sigma^{\ell^\leftarrow p^\uparrow \to \ell^\prime h X}}=
\frac{d^6\sigma^{\ell (S_\ell) + p(S_T) \to \ell^\prime h X} -
      d^6\sigma^{\ell (-S_\ell) + p(S_T) \to \ell^\prime h X}}
     {d^6\sigma^{\ell (S_\ell) + p(S_T) \to \ell^\prime h X} +
      d^6\sigma^{\ell (-S_\ell) + p(S_T) \to \ell^\prime h X}}\,\cdot
\ee
For the numerator we have
\bea
&&\frac{d\sigma^{\ell (S_\ell) + p(S_T) \to \ell^\prime h X} - 
d\sigma^{\ell (-S_\ell) + p(S_T) \to \ell^\prime h X}}
{d\xb \, dQ^2 \, dz_h \, d^2\bfP_T \, d\phi_S} = \nonumber \\
&&
\frac{2 \alpha^2}{Q^4} \left\{[1-(1-y)^2]\cos(\phi_h-\phi_S)
F_{LT}^{\cos(\phi_h-\phi_S)} + 2y\sqrt{1-y}\Big[\cos\phi_S F_{LT}^{\cos\phi_S} +\cos(2\phi_h-\phi_S)F_{LT}^{\cos(2\phi_h-\phi_S)}\Big]
\right\}. \nonumber \\ \label{l-numLT}
\eea
The denominator differs from that given in Eq.~(\ref{den1}), as it acquires
several additional terms, which also appear in the numerator of $A_{UT}$:
\bea
&&\frac{d\sigma^{\ell (S_\ell) + p(S_T) \to \ell^\prime h X} + 
d\sigma^{\ell (-S_\ell) + p(S_T) \to \ell^\prime h X}}
{d\xb \, dQ^2 \, dz_h \, d^2 \bfP_T \, d\phi_S} = \nonumber \\
&=&
\frac{2 \alpha^2}{Q^4} \biggl\{ \left[ 1+(1-y)^2 \right] \,
[F_{UU} +   \sin(\phi_h-\phi_S) F_{UT}^{\sin(\phi_h-\phi_S)}] \biggr. \nonumber \\
&& \hspace*{0.9cm} \biggl. +\> 2(1-y)[\cos 2\phi_h \, F_{UU}^{\cos 2\phi_h} +
\sin(\phi_h+\phi_S) \, F_{UT}^{\sin(\phi_h+\phi_S)} +\sin(3\phi_h-\phi_S) \, 
F_{UT}^{\sin(3\phi_h-\phi_S)}  ] \biggr. \nonumber \\
&& \hspace*{0.9cm} \biggl.
+\>  2(2-y)\sqrt{1-y} \,
[\cos\phi_h \, F_{UU}^{\cos\phi_h} +\sin \phi_S \, 
F_{UT}^{\sin \phi_S} + \sin(2\phi_h-\phi_S) \, F_{UT}^{\sin(2\phi_h-\phi_S)} ]
\biggr\} \>. 
\eea

\subsection{\label{other A} Other asymmetries}

All the other single and double spin asymmetries are either zero or related to
those already shown above. In particular, all the single spin asymmetries generated
by the lepton polarization vanish: $A_{LU} = 0$ as $F_{LU}=0$ to leading order in
$\kt/Q$ and $A_{TU} = 0$ as we have no access to the transverse polarization of the 
lepton and therefore there are no terms proportional to either $P_x^\ell$ or
$P_y^\ell$ in Eqs.~(\ref{gen-xs}) or (\ref{UT}). For the same reason we have
$A_{TT} = A_{UT}$ and $A_{TL}=A_{UL}$.

\section{\label{phen} Phenomenology of spin asymmetries }

To leading order in $(\kt/Q)$, all terms contributing to the polarized SIDIS
cross section and to the spin asymmetries can be integrated analytically,
provided we adopt a simple $\kt$ and $\pp$ dependence for the distribution and
fragmentation functions. As usual, we assume the $x$ and $\kt$ dependences
to be factorized and we assign the $\kt$ dependence a Gaussian distribution
with one free parameter to fix the Gaussian width. For the unpolarized and
helicity distribution functions and for the fragmentation function we simply use
\bea
&&f_{q/p} (x,\kt)= f_{q/p} (x)\,\frac{e^{-\kt^2/\avk}}{\pi\avk}
\label{unp-dist}\\
&&\Delta f^q_{s_z/S_L} (x,\kt)= \Delta f^q_{s_z/S_L} (x)\,\frac{e^{-\kt^2/\avk \Lsub}}
{\pi\avk \Lsub} \label{hel-dist}\\
&&D_{h/q}(z,\pp)=D_{h/q}(z)\,\frac{e^{-\pp^2/\avp}}{\pi\avp}\,,
\label{unp-frag}
\eea
where $f_{q/p}(x)$, $\Delta f^q_{s_z/S_L} (x)$ and $D_{h/q}(z)$ can be taken
from the available fits of the world data. In general, we allow for
different widths of the Gaussians for the different distributions, but take
them to be constant and flavor independent. For the Sivers and Boer-Mulders
functions, we assume a similar parametrization, with an extra multiplicative
factor $\kt$ to give them the appropriate behavior in the small $\kt$ region
\cite{Anselmino:2008sga}:
\bea
\Delta f_{q/S_T} (x,\kt) &=& \Delta f_{q/S_T}(x)\;
\sqrt{2e}\,\frac{\kt}{M\S} \; e^{-\kt^2/M^2\S}\,
\frac{e^{-\kt^2/\avk}}{\pi\avk} \nonumber \\
&=& \Delta f_{q/S_T}(x) \; \sqrt{2e}\,\frac{\kt}{M\S} \;
\frac{e^{-\kt^2/\avk\S}}{\pi\avk}
\label{Siv-dist}\\
\Delta  f^q_{s_y/p}(x,\kt) &=& \Delta  f^q_{s_y/p}(x)\;
\sqrt{2e}\,\frac{\kt}{M\BM} \; e^{-\kt^2/M^2\BM}\,
\frac{e^{-\kt^2/\avk}}{\pi\avk} \nonumber \\
&=& \Delta  f^q_{s_y/p}(x)\;
\sqrt{2e}\,\frac{\kt}{M\BM} \; \frac{e^{-\kt^2/\avk \BM}}{\pi\avk}
\,,\label{BM-dist}
\eea
where the $x$-dependent functions $\Delta f_{q/S_T}(x)$ and
$\Delta f^q_{s_y/p}(x)$ are not known, and should be determined
phenomenologically by fitting the available data on azimuthal asymmetries and
moments; the $\kt$ dependent Gaussians have been assigned a reduced width
to make sure they fulfill the appropriate positivity bounds:
\bea
&&\avk \S = \frac{\avk \, M^2\S}{\avk +  M^2 \S} \\
&&\avk \BM= \frac{\avk \, M^2\BM}{\avk + M^2 \BM}\, \cdot
\eea
Similarly, for the distribution of longitudinally polarized quarks inside a
transversely polarized proton, $\Delta  f^q_{s_z/S_T}$, and of transversely
polarized quarks inside a longitudinally polarized proton, $\Delta f^q_{s_x/S_L}$,
we set:
\bea
\Delta f^q_{s_z/S_T}  (x,\kt) &=& \Delta f^q_{s_z/S_T} (x)\;
\sqrt{2e}\,\frac{\kt}{M\LT} \; e^{-\kt^2/M\LT ^2}\,
\frac{e^{-\kt^2/\avk}}{\pi\avk} \nonumber \\
&=& \Delta f^q_{s_z/S_T}(x)\;
\sqrt{2e}\,\frac{\kt}{M\LT} \; \frac{e^{-\kt^2/\avk\LT}}{\pi\avk}
\label{g1T-dist}\\
\Delta f^q_{s_x/S_L}  (x,\kt) &=& \Delta f^q_{s_x/S_L} (x)\;
\sqrt{2e}\,\frac{\kt}{M\TL} \; e^{-\kt^2/M\TL ^2}\,
\frac{e^{-\kt^2/\avk}}{\pi\avk} \nonumber \\
&=& \Delta  f^q_{s_x/S_L}(x)\;
\sqrt{2e}\,\frac{\kt}{M\TL} \; \frac{e^{-\kt^2/\avk\TL}}{\pi\avk}
\,, \label{h1L-dist}
\eea
with
\bea
&&\avk \LT = \frac{\avk \, M^2\LT}{\avk +M^2\LT} \\
&&\avk \TL = \frac{\avk \, M^2\TL}{\avk +M^2\TL}\,\cdot
\eea
For the transversity distribution function, it is most convenient to parametrize
the following combinations
\bea
\frac{1}{2} \, \Big(\Delta f^q_{s_x/S_T}(x,\kt) + \Delta^- f^q_{s_y/S_T}(x,\kt)\Big) &=&
h_1(x,\kt) =
\, h_1(x) \, \frac{e^{-\kt^2/\avk \T}}{\pi\avk \T} \\
\frac{1}{2} \, \Big(\Delta f^q_{s_x/S_T}(x,\kt) - \Delta^- f^q_{s_y/S_T}(x,\kt)\Big) &=&
\frac{\kt ^2}{2M^2\TT} \, h^\perp_{1T}(x,\kt)  \label{T-dist} =
\, h^{\perp}_{1T}(x) \, \frac{e\,\kt^2}{M^2\TT}\,e^{-\kt^2/M\TT ^2}\, 
\frac{e^{-\kt^2/\avk}}{\pi\avk} \nonumber \\ &=&
\, h^{\perp}_{1T}(x) \, \frac{e\,\kt^2}{M^2\TT}\,
\frac{e^{-\kt^2/\avk \TT}}{\pi\avk}\,,
\eea
as these are the quantities which appear in the polarized cross section and
in the spin asymmetries. Notice that for $h_1(x,\kt)$ and $h^\perp_{1T}(x,\kt)$,
as for each of the other TMDs, we introduce their own reduced Gaussian widths
\be
\avk\T \quad\quad\quad \avk\TT=\frac{\avk M^2\TT}{\avk +M^2\TT}
\, \cdot
\ee

Finally, for the Collins fragmentation function we choose
\bea
\Delta^N  D_{h/q^\uparrow}(z,\pp) &=& \Delta^N  D_{h/q^\uparrow}(z)\;
\sqrt{2e}\,\frac{\pp}{M_h} \; e^{-\pp^2/M_h^2}\,
\frac{e^{-\pp^2/\avp}}{\pi\avp} \nonumber \\
&=&\Delta^N  D_{h/q^\uparrow}(z)\;
\sqrt{2e}\,\frac{\pp}{M_h} \;
\frac{e^{-\pp^2/\avp \C}}{\pi\avp}\,,\label{Coll-frag}
\eea
having defined
\be
\avp \C=\frac{\avp \, M_h^2}{\avp +M_h^2}\,\cdot \label{Coll-frag2}
\ee

Using the parametrizations in Eqs.~(\ref{unp-dist}-\ref{Coll-frag2}) we can
perform the $\bfk_\perp$ integrations analytically in
Eqs.~(\ref{FUU}-\ref{FUTsin(phi-2phiS)}), and re-express all the $F$ structure
functions in terms of the Gaussian parameters:
\bea
F_{UU} & = & \sum_{q} \, e_q^2 \,f_{q/p}(\xb)\,D_{h/q}(z_h)
\frac{e^{-P_T^2/\avPT}}{\pi\avPT} \label{G-FUU}\\
F_{UU}^{\cos2\phi_h} & = & -P_T^2 \, \sum_{q} \, e_q^2 \,
\frac{ \Delta f^{q}_{s_y/p}(\xb)}{M \BM}\,
\frac{\Delta^N  D_{h/q^\uparrow}(z_h)}{M_h} \,
\frac{e^{1-P_T^2/\avPT \BM}}{\pi\avPT ^3 \BM}
\, \frac{z_h\,\avk ^2 \BM \avp ^2 \C}{\avk \avp }
\label{g-FUUcos2phi} \\
F_{UU}^{\cos\phi_h} & = & -2\,\frac{P_T}{Q}\, \sum_{q} \, e_q^2 \,
f_{q/p}(\xb) \, D_{h/q}(z_h) \, \frac{e^{-P_T^2/\avPT}}{\pi\avPT^2}\, z_h \avk\,
\nonumber \\
&& +  2\,\frac{P_T}{Q}\,\sum_{q} \, e_q^2 \,
\frac{ \Delta f^{q}_{s_y/p}(\xb)}{M \BM}\, \frac{\Delta^N  D_{h/q^\uparrow}(z_h)}
{M_h} \, \frac{e^{1-P_T^2/\avPT \BM}}{\pi\avPT ^4 \BM} \,
\label{g-FUUcosphi} \\
&& \hspace{1.5cm} \times
\frac{\avk^2 \BM \avp ^2 \C}{\avk \avp} \Big[z_h^2 \avk \BM \Big(P_T^2 -
\avPT \BM \Big) + \avp \C \avPT \BM \Big]
\nonumber \\
F_{UL}^{\sin 2\phi_h} & = & P_T^2\,\sum_{q} \, e_q^2 \,
\frac{\Delta f^{q}_{s_x/S_L}(\xb)}{M \TL}\,
\frac{\Delta^N  D_{h/q^\uparrow}(z_h)}{M_h} \,
\frac{e^{1-P_T^2/\avPT \TL}}{\pi \avPT ^3 \TL} \,
\frac{z_h \avk ^2 \TL \avp ^2 \C }{\avk \avp}
\label{g-FULsin2phi} \\
F_{UL}^{\sin \phi_h} & = & -2\,\frac{P_T}{Q}\,\sum_{q} \, e_q^2 \,
\frac{\Delta f^{q}_{s_x/S_L}(\xb)}{M \TL}\,
\frac{\Delta^N  D_{h/q^\uparrow}(z_h)}{M_h} \,
\frac{e^{1-P_T^2/\avPT \TL}}{\pi \avPT ^4 \TL} \, \nonumber \\
&& \hspace{1.5cm} \times
\frac{\avk^2 \TL \avp ^2 \C}{\avk \avp}  \Big[z_h^2 \avk \TL \Big(P_T^2 -
\avPT \TL \Big) + \avp \C \avPT \TL\Big]
\label{g-FULsinphi} \\
F_{LU}^{\sin \phi_h} & = &  0 \hspace{.4cm} {\rm at \;\;leading\;\; twist}
\label{g-FLUsinphi}\\
F_{LL} & = & \sum_{q} \, e_q^2  \Delta f^{q}_{s_z/S_L}(\xb) \, D_{h/q}(z_h) \,
\frac{e^{-P_T^2/\avPT \Lsub}}{\pi\avPT \Lsub}
\label{g-FLL}\\
F_{LL}^{\cos \phi_h} & = & -2\,\frac{P_T}{Q}\,\sum_{q} \,
e_q^2\Delta f^{q}_{s_z/S_L}(\xb) \, D_{h/q}(z_h) \,
\frac{e^{-P_T^2/\avPT \Lsub}}{\pi\avPT^2 \Lsub}\, z_h  \avk \Lsub \label{g-FLLcosphi} \\
F_{UT}^{\sin(\phi_h-\phi_S)} & = &  \frac{P_T}{\sqrt{2}} \sum_{q} \, e_q^2 \, 
\frac{\Delta f_{q/S_T} (\xb)}{M\S } D_{h/q}(z_h)
\frac{e^{1/2-P_T^2/\avPT \S}}{\pi \avPT ^2 \S} \, \frac{z_h \avk ^2 \S}{\avk}\,
\label{g-FUTsin(phi-phiS)}\\
F_{LT}^{\cos(\phi_h-\phi_S)} & = &P_T \,\sum_{q} \, e_q^2 \,
\frac{\Delta f^{q}_{s_z/S_T} (\xb)}{M\LT } \, D_{h/q}(z_h)
\frac{e^{-P_T^2/\avPT \LT}}{\pi \avPT^2 \LT} \, \frac{z_h \avk ^2 \LT}{\avk}
\label{g-FUTcos(phi-phiS)} \\
F_{LT}^{\cos \phi_S} & = &   -\frac{1}{Q}\,\sum_{q} \, e_q^2 \,
\frac{\Delta f^{q}_{s_z/S_T} (\xb)}{M\LT } \, D_{h/q}(z_h) \,
\frac{e^{-P_T^2/\avPT \LT}}{\pi \avPT^3\LT} \, \frac{\avk ^2
\LT [\avp \avPT \LT + z_h^2 P_T^2 \avk \LT]}{ \avk}
\label{g-FLTcosphiS}\\
F_{LT}^{\cos(2\phi_h-\phi_S)} & = & -\frac{ P_T^2}{Q}\sum_{q}\, e_q^2 \,
\frac{\Delta f^{q}_{s_z/S_T} (\xb)}{M\LT } \, D_{h/q}(z_h) \,
\frac{e^{-P_T^2/\avPT \LT}}{\pi \avPT^3\LT} \,
\frac{z_h^2 \avk^3 \LT}{\avk}
\label{g-FLTcos2phih-phis} \\
F_{UT}^{\sin(\phi_h+\phi_S)} & = & \frac{P_T}{\sqrt{2}} \, \sum_{q}  \, e_q^2\,
h_1(\xb)\, \frac{\Delta^N  D_{h/q^\uparrow}(z_h)}{M_h} \,
\frac{e^{1/2-P_T^2/\avPT \T}}{\pi \avPT ^2 \T} \, \frac{\avp ^2 \C}{\avp}
\label{g-FUTsin(phi+phiS)}\\
F_{UT}^{\sin(3\phi_h-\phi_S)} & = & \frac{P_T^3}{\sqrt{2}}\, \sum_{q}  \, e_q^2 \,
\frac{h_{1T}^{\perp}(\xb)}{M_{TT}^2}\,\frac{\Delta^N  D_{h/q^\uparrow}(z_h)}{M_h} \,
\frac{e^{3/2-P_T^2/\avPT \TT}}{\pi \avPT ^4 \TT} \,  \frac{z_h^2 \avk^3
\TT \avp ^2 \C}{ \avk \avp}
\label{g-FUTsin(3phi-phiS)}\\
F_{UT}^{\sin\phi_S} & = &  \sqrt{2} \frac{1}{Q}\,\sum_{q}  \, e_q^2
h_1(\xb)\,\frac{\Delta^N  D_{h/q^\uparrow}(z_h)}{M_h}\,
\frac{e^{1/2-P_T^2/\avPT \T}}{\pi \avPT ^3 \T} \,
 \frac{z_h \avk \avp ^2 \C (\avPT \T - P_T^2)}{ \avp} \nonumber \\
& +&  \frac{1}{\sqrt{2}} \frac{1}{Q}\,\sum_{q} \, e_q^2 \,
\frac{\Delta f_{q/S_T} (\xb)}{M\S} D_{h/q}(z_h)
\frac{e^{1/2-P_T^2/\avPT \S}}{\pi \avPT ^3 \S} \,   \frac{\avk ^2 \S
(\avp \avPT \S + z_h^2 P_T^2 \avk \S)}{\avk}\,
\label{g-FUTsinphiS}\\
F_{UT}^{\sin(2\phi_h-\phi_S)}  & = & -\sqrt{2}\,\frac{P_T^2}{Q}\sum_{q} \, e_q^2\,
\frac{h_{1T}^{\perp}(\xb)}{M_{TT}^2}\,\frac{\Delta^N  D_{h/q^\uparrow}(z_h)}{M_h} \,
\nonumber \\ & & \times \,
\frac{e^{3/2-P_T^2/\avPT \TT}}{\pi \avPT ^5 \TT} \,
\frac{z_h \avk^3 \TT \avp ^2 \C \Big[z_h^2 \avk \TT (P_T^2 -\avPT \TT) +
2 \avp \C \avPT \TT\Big] }{\avk \avp}
\nonumber \\ & - &
\frac{1}{\sqrt{2}} \,\frac{P_T^2}{Q}\,\sum_{q} \, e_q^2\,
\frac{\Delta f_{q/S_T} (\xb)}{M\S} D_{h/q}(z_h)
\frac{e^{1/2-P_T^2/\avPT \S}}{\pi \avPT ^3 \S} \, \frac{z_h^2  \avk ^3 \S}{\avk}
\label{g-FUTsin(phi-2phiS)}
\eea
where
\bea
\avPT &=& \avp + z_h^2 \, \avk \nonumber \\
\avPT_{I} &=& \avp + z_h^2 \, \avk_{I} \quad (I = S, L, LT) \label{PTI} \\
\avPT_{J} &=& \avp \C + z_h^2 \, \avk_{J} \quad (J = T, BM, TL,TT)\>. \nonumber
\eea

The unpolarized SIDIS cross section and all the asymmetries presented in
Section~\ref{asy} can now be rewritten in terms of the Gaussian-integrated
$F$'s, which depend on the TMDs. In order to single out information on a
particular TMD from the measurements of the asymmetries, one has to disentangle
the different azimuthal dependences. For example, the unpolarized cross section,
see Eq.~(\ref{den1-s}), includes the usual unpolarized collinear
SIDIS cross section, the Cahn effect proportional to $\cos\phi_h$ (studied in
Ref.~\cite{Anselmino:2005nn}), and a contribution generated by a combined
Boer-Mulders and Collins effect, which appears in terms proportional to
$\cos2\phi_h$ and $\cos\phi_h$. Similarly, in the numerator of the $A_{UT}$
single spin asymmetry, Eq.~(\ref{num1-s}), the Sivers and Collins effects are
both simultaneously at work, together with other azimuthal modulations.
To extract single effects, one introduces appropriate azimuthal moments of the asymmetries,
defined as
\be
A_{S_\ell S}^{W(\phi_h,\phi_S)} \equiv 
2\,\frac{\displaystyle \int d\phi_h \,d\phi_S \,
[d\sigma^{\ell(S_\ell) + p(S)\to \ell^\prime h X} -
d\sigma^{\ell(S_\ell) + p(-S) \to \ell^\prime h X}]\,W(\phi_h,\phi_S)}
{\displaystyle \int d\phi_h\, d\phi_S \,
[d\sigma^{\ell(S_\ell) + p(S) \to \ell^\prime h X} +
d\sigma^{\ell(S_\ell) + p(-S) \to \ell^\prime h X}]}\,,
\label{asy-mom}
\ee
where the function $W(\phi_h,\phi_S)$ is an appropriate ``weighting phase'' which, upon integration,
singles out one individual term of the asymmetry. For instance, to isolate the Sivers effect one can consider the
$\sin(\phi_h-\phi_S)$ azimuthal moment of the $A_{UT}$ asymmetry:
\be
A_{UT}^{\sin(\phi_h-\phi_S)}=2\,
\frac{\displaystyle\int d\phi_h \,d\phi_S \,
[d\sigma^{\ell\, p^\uparrow \to \ell^\prime h X} -
d\sigma^{\ell\, p^\downarrow \to \ell^\prime h X}]\,\sin(\phi_h-\phi_S)}
{\displaystyle\int d\phi_h\, d\phi_S \,
[d\sigma^{\ell \,p^\uparrow \to \ell^\prime h X} +
d\sigma^{\ell\, p^\downarrow \to \ell^\prime h X}]}\,\cdot
\ee
The $W$ weight selects the Sivers term of the asymmetry in the numerator, while the
integration over the azimuthal angles $\phi_S$ and $\phi_h$ leaves only the
first term of the unpolarized cross section, Eq.~(\ref{den1-s}), in the
denominator: thus, this azimuthal moment is simply proportional to the ratio
$\int F_{UT}^{\sin(\phi_h-\phi_S)} / \int F_{UU}$.

Furthermore, experimental data deliver these azimuthal moments as a function
of one variable at a time, either $\xb$, $z_h$ or $P_T$. Therefore, one has
to integrate the numerator and denominator {\it separately} over all variables
but one, in order to obtain the appropriate expression to be compared with the
data. Clearly, no simplification of common terms in the numerator and denominator
can be made before the integrations have been performed (notice also that $y$
is a function of both $\xb$ and $Q^2$).

Let us consider, as an explicit example, the Sivers azimuthal moment
$A_{UT}^{\sin(\phi_h-\phi_S)}(z_h)$, as function of $z_h$ alone. Using the
Gaussian-integrated expression of $F_{UT}^{\sin(\phi_h-\phi_S)}$ of
Eq.~(\ref{g-FUTsin(phi-phiS)}) and integrating analytically over $\bfP_T$
we obtain
\be
A_{UT}^{\sin(\phi_h-\phi_S)}(z_h)= \,A\S\,
\frac{\displaystyle \int d\xb \, dQ^2  \, \frac{1+(1-y)^2}{Q^4}
\sum _q e_q^2 \,\Delta ^N f_{q/S_T}(\xb)\,D_{h/q}(z_h)}
{\displaystyle \int d\xb \, dQ^2 \, \frac{1+(1-y)^2}{Q^4}
\sum _q  e_q^2\, f_{q/p}(\xb)\,D_{h/q}(z_h)}\,,
\label{siver-asym}
\ee
where $A\S$ is a factor which only depends on $z_h$ and on the free parameters
which give the Gaussian widths for the distribution and fragmentation functions
\be
A\S=\frac{z_h}{4\,M\S}\,\sqrt{\frac{2\,e\,\pi}{\avPT \S}}\,\frac{\avk \S ^2}
{\avk}\,\cdot
\ee
Notice the further dependence on $z_h$ hidden in $\avPT \S$, Eq.~(\ref{PTI}).


Repeating similar procedures one can extract information on the other TMDs.
The azimuthal moment $A_{UT}^{\sin(\phi_h+\phi_S)}$, obtained using the
weighting phase $W(\phi_h,\phi_S)=\sin(\phi_h+\phi_S)$ in Eq.~(\ref{asy-mom}) 
with unpolarized leptons, selects the Collins effect,
coupled to the transversity distribution $F^{+-}_{+-}(x)=\Delta _T q(x)=h_1(x)$.
In this case, the azimuthal moment is sensitive to the ratio
$F_{UT}^{\sin(\phi_h + \phi_S)} / F_{UU}$, and precisely:
\be
A_{UT}^{\sin(\phi_h+\phi_S)}(z_h)= A\C\,
\frac{\displaystyle \int d\xb \, dQ^2 \, \frac{2(1-y)}{Q^4}\;
\sum _q e_q^2 \,h_1(\xb)\,\Delta^N D_{h/q^\uparrow}(z_h)}
{\displaystyle\int d\xb \, dQ^2 \, \frac{1+(1-y)^2}{Q^4}
\sum _q e_q^2\, f_{q/p}(\xb)\,D_{h/q}(z_h)}\,,
\ee
with
\be
A\C=\frac{1}{4 \, M_h}\sqrt{\frac{2 \, e \, \pi}{\avPT \C}}\,
\frac{\avp \C ^2}{\avp}\;\cdot
\ee

One can further exploit the $A_{UT}$ asymmetry, to isolate and measure the
transverse distribution function $ F^{-+}_{+-}(x)=h_{1T}^\perp(x)$, by
weighting the single spin asymmetry numerator with the phase
$W(\phi_h,\phi_S)=\sin(3\phi_h-\phi_S)$, obtaining:
\be
A_{UT}^{\sin(3\phi_h-\phi_S)}(z_h)= A\TT \,
\frac{\displaystyle \int d\xb \, dQ^2 \, \frac{2(1-y)}{Q^4}\;
\sum _q e_q^2 \,h_{1T}^\perp(\xb)\,\Delta^N D_{h/q^\uparrow}(z_h)}
{\displaystyle\int d\xb \, dQ^2 \, \frac{1+(1-y)^2}{Q^4}
\sum _q e_q^2\, f_{q/p}(\xb)\,D_{h/q}(z_h)}\,,
\ee
where
\be
A\TT=\frac{3 \, e \, z_h^2}{8 \,M\TT ^2 \,M_h \, \avPT \TT}\sqrt{\frac{2 \, e \, \pi}
{\avPT \TT}}\,\frac{\avk \TT ^3}{\avk}\,\frac{\avp \C ^2}{\avp}\;\cdot
\ee


One can write similar expressions for all other asymmetries, which we do not
report here. From $A_{UL}^{\sin\phi_h}$ and $A_{UL}^{\sin2\phi_h}$ one can
obtain information on $\Delta f _{s_x/S_L}$, while $A_{LT}^{\cos\phi_S}$,
$A_{LT}^{\cos(\phi_h-\phi_S)}$ and $A_{LT}^{\cos(2\phi_h-\phi_S)}$ depend on
$\Delta f _{s_z/S_T}$.
$A_{UT}^{\sin\phi_S}$ and $A_{UT}^{\sin(2\phi_h-\phi_S)}$ are more
complicated to analyze as they receive contributions from the Sivers
distribution function (both of them) and, in addition, from the transversity
distribution $h_1(x)$ ($A_{UT}^{\sin\phi_S}$) and from $h_{1T}^\perp$
($A_{UT}^{\sin(2\phi_h-\phi_S)}$).

Let us consider in more details the unpolarized cross section, to which,
remarkably, a similar ``weighting'' procedure can be applied. In fact,
one can introduce the average value of $W(\phi_h)$ with an expression similar to Eq.~(\ref{asy-mom}) in which
the unpolarized cross section appears in the numerator as well as in the denominator
\be
\langle W(\phi_h)\rangle=
\frac{\displaystyle \int d\phi_h \,d\phi_S \,
[d\sigma^{\ell p^\uparrow \to \ell^\prime h X} +
d\sigma^{\ell p^\downarrow \to \ell^\prime h X}]\,W(\phi_h)}
{\displaystyle \int d\phi_h\, d\phi_S \,
[d\sigma^{\ell p^\uparrow \to \ell^\prime h X} +
d\sigma^{\ell p^\downarrow \to \ell^\prime h X}]}\,\cdot
\label{val-med}
\ee
For instance,
weighting the unpolarized cross section with $W(\phi_h)=\cos 2\phi_h$ one can gain
direct access to the Boer-Mulders function, coupled to the Collins function
(on which independent information can be obtained):
\be
 \langle \cos 2\phi_h \rangle=
A\BM \,
\frac{\displaystyle \int d\xb dQ^2 \, \frac{(1-y)}{Q^4}\sum _q e_q^2 \,
\Delta f^q_{s_y/p}(\xb)\,\Delta^N D_{h/q^\uparrow}(z_h)}
{\displaystyle \int d\xb dQ^2 \, \frac{1+(1-y)^2}{Q^4} \sum _q e_q^2\,
f_{q/p}(\xb)\,D_{h/q}(z_h)}\,,
\ee
with
\be
A\BM=-\frac{e\,z_h}{M\BM M_h \, \avPT \BM}\frac{\avk \BM ^2}{\avk}\,
\frac{\avp \C ^2}{\avp}\;\cdot
\ee
Analogously, using $W(\phi_h)=\cos\phi_h$, one has
\be
\langle \cos \phi_h \rangle  =
\frac{\displaystyle \int d\xb \, dQ^2 \, \frac{(2-y)\sqrt{1-y}}{Q^4}
\sum _q e_q^2 \, \Big[A_{\rm unp}\,f_{q/p}(\xb)\,D_{h/q}(z_h) +
B\BM \Delta f^q_{s_y/p}(\xb)\,\Delta^N D_{h/q^\uparrow}(z_h)\Big]}
{\displaystyle \int d\xb \, dQ^2 \, \frac{1+(1-y)^2}{Q^4}
\sum _q e_q^2\, f_{q/p}(\xb)\,D_{h/q}(z_h)}
\ee
with
\be
A_{\rm unp}=-\,z_h\,\frac{\avk}{Q}\,\sqrt{\frac{\pi}{\avPT}}\,\;,\;\;\;\;\;\;
B\BM  = \frac{e\sqrt{\pi}}{2\,Q \,M\BM \,M_h}
\frac{\avk \BM ^2}{\avk}\,\frac{\avp \C ^2}{\avp}\,\frac{[\avp \C + \avPT \BM]}{\avPT \BM ^{3/2}}\;\cdot
\ee


\newpage

\section{\label{conc} Conclusions and further remarks}

The study of the 3-dimensional structure of protons and neutrons is one of
the central issues in hadron physics, with many dedicated experiments, either
running (COMPASS at CERN, CLASS at JLab, STAR and PHENIX at RHIC), approved 
(JLab upgrade) or being planned (ENC/EIC Colliders). The transverse momentum 
dependent partonic distribution and fragmentation functions, together with the 
generalized parton distributions, play a crucial role in gathering and interpreting
information towards a true 3-dimensional imaging of the nucleons.
TMDs can be accessed in several experiments, but the main source of information
is semi-inclusive deep inelastic scattering of leptons off polarized
nucleons. The theoretical framework in which the experimental information is
analyzed is the QCD factorization scheme.

We have used here an intuitive approach to TMD factorization in SIDIS and
shown that one can re-derive, at leading order, the most general expression
of the polarized cross section, obtained within the QCD factorization scheme
by other authors~\cite{Mulders:1995dh, Kotzinian:1994dv, Bacchetta:2006tn}.
All azimuthal dependences are
precisely generated by the properties of the helicity amplitudes, which we use
to describe the factorized steps of the process: the partonic distributions,
the elementary interaction and the quark fragmentation.

We have obtained explicit expressions for all the SIDIS spin asymmetries
and the cross section azimuthal dependences which allow to extract
information on the TMDs. Indeed, some of them have already been used to study
the Sivers~\cite{Sivers:1989cc, Sivers:1990fh}, the Cahn~\cite{Cahn:1978se,
Cahn:1989yf} and the Collins~\cite{Collins:1992kk} effects. Simplified
expressions, based on a Gaussian $\kt$ and $p_\perp$ dependence of the
distribution and fragmentation functions, recently supported by data
\cite{Schweitzer:2010tt}, have been given; they might be useful for fast and
simple analyses of the experimental data.

We wonder, at this stage, whether the same approach can be used for other
processes. It works, with the same validity as for SIDIS, for Drell-Yan
processes (D-Y) \cite{Anselmino:2009zza}, where our helicity amplitudes for
the different factorized steps reproduce the most general azimuthal structure
of the cross section as obtained in the TMD factorization \cite{Arnold:2008kf}.
As commented in the Introduction, both in SIDIS and D-Y the presence of two
different natural scales, a small and a large one, is crucial for the validity
of the QCD TMD factorization.

Our approach was actually first introduced for processes with a single large
scale, like $p \, p \to \pi X$, with large $P_T$ pions \cite{Anselmino:2005sh}.
These are the processes for which the largest single spin asymmetries 
have been observed and might be generated by TMDs \cite{Anselmino:1994tv,
Anselmino:1998yz, D'Alesio:2004up}. However, TMD factorization has not been
proven in these cases. Despite that, an extension of the intuitive approach
used for SIDIS -- {\it and shown to be perfectly equivalent to the QCD TMD
factorization scheme} -- is natural. That was the guiding idea in
Ref. \cite{Anselmino:2005sh}; each proton ``emits" a parton, the two partons
interacts and one of the final parton fragments into the observed hadron.
All intrinsic motions are taken into account and phases appear in the helicity
amplitudes. The difference with SIDIS processes is that, in this case, the
measured large $P_T$ of the final hadron is generated by the hard elementary
scattering, and all intrinsic motions are integrated over. As a consequence,
the phase integrations strongly suppress the relevance of most TMDs, with the
exception of the Sivers and Collins effects~\cite{Anselmino:2008uy,
Anselmino:2009hk}, which combine into the observed asymmetry, and cannot be
separated unless one could resolve the internal structure of the final
jet \cite{D'Alesio:2010am}.

A global simultaneous phenomenological analysis of single spin asymmetries in SIDIS and $pp$
interactions is, at the moment, rather difficult. Apart from the validity
of the factorization scheme in both cases, another important open point is
the universality of the Sivers functions; it is not clear whether or not they
should be the same in the two processes or should be corrected by some gauge
color factors \cite{Gamberg:2010tj}. In any case it is worth trying to explore
the possibility to have a unique description of SSAs in different processes,
based on TMDs; work in this direction is in progress and will be presented elsewhere.

\section{Acknowledgements}
We are grateful to Aram Kotzinian for several useful discussions.
We acknowledge support of the European Community - Research Infrastructure
Activity under the FP7 ``Structuring the European Research Area''
program (HadronPhysics2, Grant agreement 227431).
We acknowledge partial support by MIUR under Cofinanziamento PRIN 2008.
This work is partially supported by the Helmholtz Association through
funds provided to the virtual institute ``Spin and strong QCD''(VH-VI-231), and by the
DOE Contract No. DE-AC05-06OR23177, under which Jefferson Science Associate, LLC, operates
the Jefferson Laboratory.

\newpage

\appendix

\section{Helicity Amplitudes}\label{helicity-appendix}
We show the explicit computation of the helicity amplitudes
$\hat M_{\lambda_3 \lambda_4; \lambda_1 \lambda_2}$ for the non-planar process
$\ell(k_1,\lambda_1) + q(k_2,\lambda_2) \rightarrow \ell ^\prime (k_3,\lambda_3)
+ q^\prime (k_4,\lambda_4)$, in the $\gamma^* - p$ c.m. frame of Fig. 1. We exploit the
spinor helicity technique, adopting the conventions of Ref. \cite{Leader:2001gr}.
At LO in QED, when neglecting all masses, there are two independent helicity
amplitudes:
\bea
\hat M_{++;++} &=&
\frac{e_q \, e^2}{\hat t} \, \langle {q^\prime}^+|\gamma^{\mu}| q^+
\rangle \, \langle {\ell ^\prime}^+|\gamma_{\mu}|\ell ^+\rangle =
\frac{e_q \, e^2}{\hat t} \, \langle 4^+|\gamma^{\mu}| 2^+
\rangle \, \langle 3^+|\gamma_{\mu}|1^+\rangle \\
\hat M_{+-;+-} &=&
\frac{e_q \, e^2}{\hat t} \, \langle {q^\prime}^-|\gamma^{\mu}| q^-
\rangle \, \langle {\ell ^\prime}^+|\gamma_{\mu}|\ell ^+\rangle =
\frac{e_q \, e^2}{\hat t} \, \langle 4^-|\gamma^{\mu}| 2^-
\rangle \, \langle 3^+|\gamma_{\mu}|1^+\rangle\,,
\eea
which can be written as
\bea
\hat M_{++;++}&=& 2\, \frac{e_q \, e^2}{\hat t}\,[43] \, \langle 1 2\rangle
\label{dixamp1}\\
\hat M_{+-;+-}&=& 2\, \frac{e_q \, e^2}{\hat t}\,[23] \, \langle 1 4\rangle
\label{dixamp2} \,,
\eea
where
\bea
\bar u_-(k_i) \, u_+(k_j) &\equiv& \langle ij\rangle =-[ij]^*
= \sqrt{k_i^+ k_j^-} e^{-i(\phi_i-\phi_j)/2} -
  \sqrt{k_i^- k_j^+} e^{i(\phi_i-\phi_j)/2} \label{dix3} \\
\bar u_+(k_i) \, u_-(k_j) &\equiv& [ij] = - \langle ij\rangle^* \>,
\label{dix4}
\eea
with $k^{\pm}=k^0\pm k^3$.

In the $\gamma^*-p$ c.m. frame we have (see Ref.~\cite{Anselmino:2005nn} for
 details):
\bea
&& k_1 = E ( 1,\sin\theta, 0, \cos\theta)\nonumber\\
&& q = \frac{1}{2} \Big( W-\frac{Q^2}{W},0,0,W+\frac{Q^2}{W} \Big) \nonumber\\
&& k_2=\Big(xP_0+\frac{k_\perp^2}{4xP_0},\bkt,-xP_0+\frac{k_\perp^2}{4xP_0}\Big) \label{k2}\\
&& k_3=k_1-q\nonumber\\
&& k_4=k_2+q\nonumber\\
&&\phi_{1,3}=0\;,\;\;\;\;\phi_{2,4}=\phi_\perp\;,\nonumber
\eea
where, neglecting the proton mass:
\bea
&& x=\frac{1}{2}\xb\left(1+\sqrt{1+4\frac{\kt^2}{Q^2}}\right)\nonumber \\
&& E= \frac{s-Q^2}{2W}=\frac{\sqrt{s}}{2}\frac{1-\xb y}{\sqrt{y(1-\xb)}}
\nonumber \\
&& Q^2=\xb y \, s \quad\quad W=\sqrt{y(1-\xb)s} \nonumber \\
&& P_0=\frac{1}{2}\Big(W+\frac{Q^2}{W}\Big)=\frac{\sqrt{s}}{2}\sqrt{\frac{y}{1-\xb}}\\
&&\frac{1}{2}\Big(W-\frac{Q^2}{W}\Big)=
\frac{\sqrt{s}}{2}\sqrt{\frac{y}{1-\xb}}(1-2\xb) \nonumber \\
&&\cos\theta=\frac{1+(y-2)\xb}{1-y\xb} \quad\quad
  \sin\theta=\frac{2\sqrt{\xb(1-\xb)(1-y)}}{1-y\xb} \,\cdot \nonumber
\eea

These relations allow us to express all the $k_i^\pm$ components in terms of
$\xb$ and $y$~\cite{Anselmino:2005nn}:
\begin{eqnarray}
k_1^{+}&=&E(1+\cos\theta)=\sqrt{s}\sqrt{\frac{1-\xb}{y}} \nonumber \\
k_1^{-}&=&E(1-\cos\theta)=\sqrt{s}\frac{\xb(1-y)}{\sqrt{y(1-\xb)}} \nonumber\\
k_3^{+}&=&E(1+\cos\theta)-W=\sqrt{s}\sqrt{\frac{1-\xb}{y}}(1-y) \nonumber \\
k_3^{-}&=& E(1-\cos\theta)-\frac{Q^2}{W}=\sqrt{s}\frac{\xb}{\sqrt{y(1-\xb)}}
\label{dix_kin}\\
k_2^{+}&=&\frac{k_{\perp}^2}{2 x P_0}=\frac{k_{\perp}^2}{x\sqrt{s}}
\sqrt{\frac{1-\xb}{y}}\nonumber\\
k_2^{-}&=&2x P_0=x \sqrt{s}\sqrt{\frac{y}{1-\xb}} \nonumber\\
k_4^{+}&=&\frac{k_{\perp}^2}{2 x P_0}+W=\sqrt{s}\sqrt{\frac{1-\xb}{y}}\left[\frac{k_{\perp}^2}{x s}+y\right] \nonumber\\
k_4^{-}&=&2x P_0-\frac{Q^2}{W}=\sqrt{s}\sqrt{\frac{y}{1-\xb}}[x-\xb] \nonumber\\
\phi_1&=&\phi_3=0\,,\qquad \phi_2=\phi_4=\phi_{\perp}\,.\nonumber
\end{eqnarray}

From Eqs.~(\ref{dixamp1})--(\ref{dix4}) we get:
\be
 \hat M_{++;++} = 2\frac{e_q e^2}{\hat t}\left[\sqrt{k_1^- k_2^{+}} -
 \sqrt{k_1^+ k_2^-} \, e^{i\phi_{\perp}}\right]
\times\left[\sqrt{k_3^- k_4^{+}} -
\sqrt{k_3^+k_4^{-}} \, e ^{-i\phi_{\perp}}\right]
\ee
\be
 \hat M_{+-;+-} = 2\frac{e_q e^2}{\hat t}\left[ \sqrt{k_2^- k_3^+} -
 \sqrt{k_2^+ k_3^{-}} \, e^{i\phi_{\perp}} \right]
 \times\left[ \sqrt{k_1^{+}k_4^-} -
\sqrt{k_1^{-} k_4^{+}} \, e ^{-i\phi_{\perp}} \right] \>.
\ee
Exploiting Eqs.~(\ref{dix_kin}) we can finally compute the amplitudes as
function of $y$, $Q^2$ and $k_{\perp}$:
\be
\hat M_{++;++} = e_q \, e^2 \left[
\frac{1}{y}\left( 1 + \sqrt{1+4\frac{\kt^2}{Q^2}} \right) e^{+i\phi_\perp} -
\frac{1-y}{y}\left( 1 - \sqrt{1+4\frac{\kt^2}{Q^2}} \right) e^{-i\phi_\perp}
- 4\ \frac{\sqrt{1-y}}{y}\,\frac{\kt}{Q}\label{M++a} \right]
\ee
\be
\hat M_{+-;+-} = e_q \, e^2 \left[
\frac{1-y}{y}\left( 1 + \sqrt{1 + 4\frac{\kt^2}{Q^2}} \right) e^{-i\phi_\perp} -
\frac{1}{y}\left( 1 - \sqrt{1 + 4\frac{\kt^2}{Q^2}} \right) e^{+i\phi_\perp}
- 4 \frac{\sqrt{1-y}}{y}\,\frac{\kt}{Q}\label{M+-a} \right] \,\cdot
\ee

\bigskip

\section{Helicity formalism and helicity transformations}\label{hel-transf}

All our analytical and numerical computations of the SIDIS cross section,
Eq.~(\ref{gen-xs}), are performed in the $\gamma^* - p$ center of mass frame
(c.m.), with the kinematics represented in Fig.~\ref{fig.1}. However, in our helicity formalism all components of the
polarization vectors (like in Eqs. (\ref{rho-q}) and (\ref{rho-p})) and of
the transverse momenta which enter the definition of the TMDs, refer to the
appropriate helicity frame of the corresponding particle. Then, in order
to perform our calculations, we have to express the helicity frame variables
in terms of the c.m.~ones, which requires some care.

For the proton, which moves along $-\hat{\bfZ}_{cm}$, the helicity frame
$(\hat{\bfX}_{p},\hat{\bfY}_{p},\hat{\bfZ}_{p})$, as reached from the
$\gamma^* - p$ c.m. frame, is given by (as discussed in Appendix~D of
Ref.~\cite{Anselmino:2005sh}):
\be
\hat{\bfX}_{p} =  \hat{\bfX}_{cm} \;\;\;\;\;\;
\hat{\bfY}_{p} = -\hat{\bfY}_{cm} \;\;\;\;\;\;
\hat{\bfZ}_{p} = -\hat{\bfZ}_{cm} \>,
\label{axes-phel-cm}
\ee
so that
\bea
&& \hat{\bfk}_\perp =
\cos \varphi_\perp \,\hat{\bfX}_{p} + \sin \varphi_\perp \,\hat{\bfY}_{p} =
\cos \phi_\perp \,\hat{\bfX}_{cm} + \sin \phi_\perp \,\hat{\bfY}_{cm} =
\cos \varphi_\perp \,\hat{\bfX}_{cm} - \sin \varphi_\perp \,\hat{\bfY}_{cm}
\nonumber \\
&& \bfk_2 = \bfk_\perp - \left( \xb P_0 - \frac{k_\perp^2}{4 \xb P_0}
\right) \hat{\bfZ}_{cm} \\
&& \bfS_T =
\cos \varphi_{S} \,\hat{\bfX}_{p} + \sin \varphi_{S} \, \hat{\bfY}_{p} =
\cos \phi_{S} \, \hat{\bfX}_{cm} + \sin \phi_{S} \, \hat{\bfY}_{cm} =
\cos \varphi_{S} \,\hat{\bfX}_{cm} - \sin \varphi_{S} \,\hat{\bfY}_{cm}
\nonumber \>,
\label{kcm}
\eea
which implies $\varphi_{\perp, S} = 2\pi - \phi_{\perp, S}$.
As long as there is no ambiguity we use $\varphi$ for angles defined in the
helicity frames and $\phi$ for angles defined in the c.m. frame, following
the notations of Fig. 1.

It is less straightforward to deal with the quark polarization vector,
$\bfP^q = (P_x^q, P_y^q, P_z^q)$, which describes intrinsic properties
of the proton constituents, and is defined in the {\it quark helicity frame}. In order
to keep the same definitions, through the helicity formalism, of the polarized
TMDs as in Ref.~\cite{Anselmino:2005sh}, we have to define $\bfP^q$ in the quark
helicity frame as reached from the proton helicity frame. The axes $\hat{\bfx}_q, \hat{\bfy}_q, \hat{\bfz}_q$ of the quark helicity frame are then given by
\cite{Leader:2001gr, Anselmino:2005sh}:
\bea
\hat{\bfz}_q &=& \hat{\bfk}_2  \label{zq} \\
\hat{\bfy}_q &=& \hat{\bfZ}_{p} \times \hat{\bfk}_\perp
                = - \hat{\bfZ}_{cm} \times \hat{\bfk}_\perp \label{yq}\\
\hat{\bfx}_q &=& \hat{\bfy}_q \times \hat{\bfz}_q =
(\hat{\bfZ}_{p} \times \hat{\bfk}_\perp) \times\hat{\bfk}_2
= -(\hat{\bfZ}_{cm} \times \hat{\bfk}_\perp) \times\hat{\bfk}_2
               \>. \label{xq}
\eea
Notice that the quark helicity frame as reached from the c.m. frame
($\hat{\bfZ}_{cm}$) is different from the quark helicity frame as reached from
its parent proton helicity frame ($\hat{\bfZ}_{p}$); although the $\hat{\bfz}_q$
axes obviously coincide, $\hat{\bfx}_q$ and $\hat{\bfy}_q$ have opposite signs,
Eqs. (\ref{xq}) and (\ref{yq}). Therefore, when referring to the kinematical
configuration of Fig.~1, which we use throughout the paper, we have to take the
$x$ and $y$ component of the quark polarization vector, $P_x^q$ and $P_y^q$,
with opposite signs with respect to those obtained from Eq.~(\ref{defF}); this
has been done in Eqs.~(\ref{fxqs}) and (\ref{fyqs}).


\bigskip

\section{Analysis of the fragmentation process}\label{phi}

\begin{figure}[b]
\begin{center}
\includegraphics[width=12.5truecm]{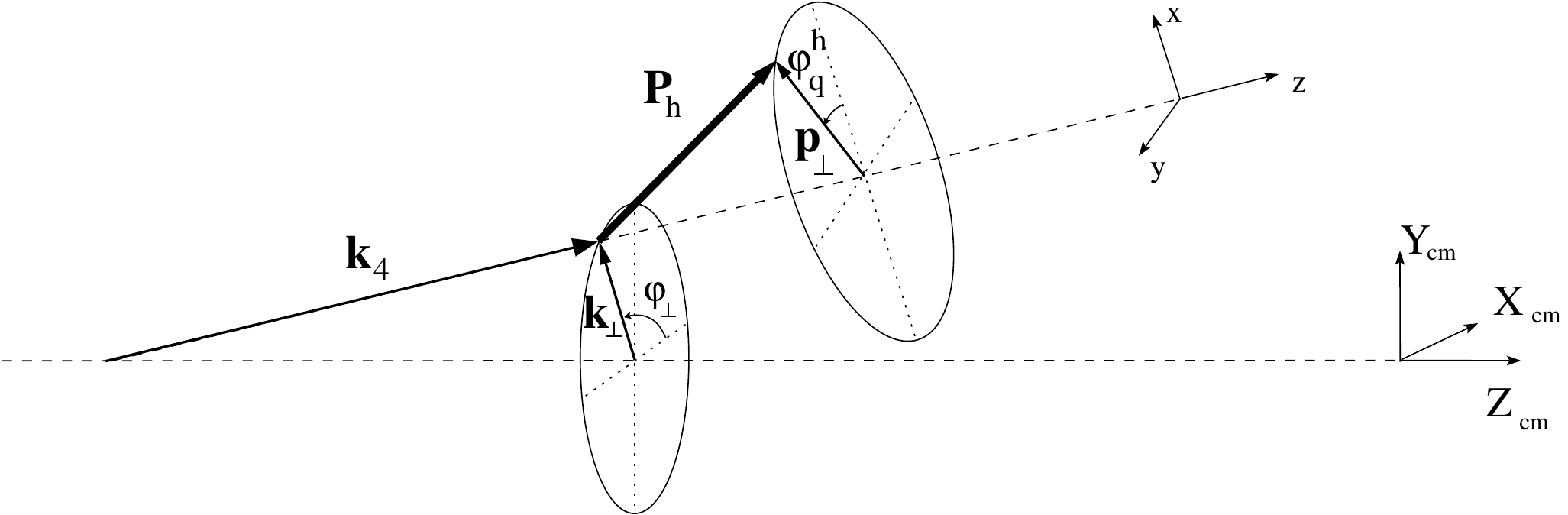}
\caption{\small Kinematics of the fragmentation process.}
\label{fig.3}
\end{center}
\end{figure}

Let us now focus on the azimuthal angle $\varphi_{q}^h$ involved in the
fragmentation process. This is the azimuthal angle of the momentum $\bfP_h$
of the final hadron around the direction $\bfk_4$ of the fragmenting quark $q$,
{\it as defined in the quark $q$ helicity frame}, see  Fig.~\ref{fig.3}. Notice that the fragmenting
quark, in the $\gamma^* - p$ c.m. frame, has a longitudinal component along the
positive $Z_{cm}$ axis. Its helicity frame, as reached from the $\gamma ^*-p$
c.m. frame, is given by Ref.~\cite{Anselmino:2005sh}:
\bea
&& \hat{\bfz} = \hat{\bfk}_4 \nonumber \\
&& \hat{\bfy} = \hat{\bfZ}_{cm} \times \hat{\bfk}_{\perp} \\
&& \hat{\bfx} =  \hat{\bfy} \times \hat{\bfz} \,, \nonumber
\eea
where $\hat{\bfk}_{\perp}$ is the unit transverse component -- with respect
to the $Z_{cm}$ direction -- of the outgoing quark, $\hat{\bfk}_{4}$.

%
%

In the quark helicity frame, $\varphi_{q}^h$ coincides with the azimuthal angle
which identifies the hadron transverse momentum $\bpp$, therefore
\bea
&&\cos \varphi_{q}^h = \hat{\bfp}_\perp \cdot \hat{\bfx} \nonumber \\
&&\sin \varphi_{q}^h = \hat{\bfp}_\perp \cdot \hat{\bfy} \,.
\eea
By using the SIDIS kinematics as reported in Ref.~\cite{Anselmino:2005nn}, one
finds
\bea
&& \cos \varphi_{q}^h = \frac{1}{p_\perp \, |\bfk _4|}
[P_T \, k_4^Z \, \cos(\phi _h - \phi_\perp) - P_h^Z \, \kt ] \nonumber \\
&& \sin \varphi_{q}^h = \frac{P_T}{p_\perp} \,
\sin(\phi _h-\phi_\perp)\,,
\label{phi-qh}
\eea
where the superscript $Z$ refers to the $\gamma^* - p$ c.m. frame, where one
measures $\bfP _h = (P_T \cos\phi_h, P_T \sin\phi_h, P_h^Z)$, and
\bea
&&P_h^Z=\frac{z_h^2 \, W^2-P_T^2}{2 \, z_h \, W} \nonumber \\
&&k_4^Z=\frac{W}{2}
        \left(\frac{1-x}{1-\xb}+\frac{\xb}{x} \frac{\kt^2}{Q^2}\right) \\
&& |\bfk _4|=\sqrt{\frac{W^2}{4}
   \left(\frac{1-x}{1-\xb}+\frac{\xb}{x} \frac{\kt^2}{Q^2}\right)^2+\kt ^2}
\>\>, \nonumber
\eea
as derived in Ref.~\cite{Anselmino:2005nn}.

At $\mathcal{O}(k_\perp/Q)$ one simply has
\bea
&& \cos \varphi_q^h =\frac{P_T}{p_\perp}
           \left[\cos(\phi _h-\phi_\perp) -z_h\frac{\kt}{P_T} \right]
\nonumber\\
&& \sin \varphi_q^h = \frac{P_T}{p_\perp} \, \sin(\phi _h-\phi_\perp)\,,
\label{phi-qh1}
\eea
having neglected terms $\mathcal{O}(k_\perp^2/W^2)$ and $\mathcal{O}(P_T^2/W^2)$.

\bigskip

\section{Tensorial Analysis}\label{t-An}

Eqs.~(\ref{FUU})-(\ref{FUTsin(phi-2phiS)}) are obtained using a simple euclidean
tensorial analysis, as outlined in what follows. In general, the tensorial
structure of each of the $F$'s functions defined in Eqs.~(\ref{FUU})-(\ref{FUTsin(phi-2phiS)}) can be reduced to a linear combination of the convolutions
\bea
T^i     &=& \int d^2\bkt \; \Delta f (x,\kt) \, \kt ^i \,
\Delta D (z,\pp) \label{Ti}\\
T^{ij}  &=& \int d^2\bkt \; \Delta f (x,\kt) \, \kt ^i \, \kt ^j \,
\Delta D (z,\pp) \label{Tij} \\
T^{ijl} &=& \int d^2\bkt \; \Delta f (x,\kt) \, \kt ^i \, \kt ^j \, \kt ^l \,
\Delta D (z,\pp) \,,\label{Tijl}
\eea
where we have denoted by $\Delta f$ ($\Delta D$) any distribution
(fragmentation) function appearing in the definition of the particular $F$
function one is considering, while the $\kt ^i$, $i=X,Y$ ($X$ and $Y$
refer to the $\gamma^*-p$ c.m. frame, we have dropped the $cm$ subscript)
are the components of the $\bkt$ transverse momentum vector,
$\kt ^X = \kt \cos\phi_{\perp}$, $\kt ^Y =
\kt \sin\phi_{\perp}$. One should bear in mind that $\pp$ is not an independent
quantity, as it can be expressed in terms of $\kt$ and  $P_T$.
Notice that $T^{i}$, $T^{ij}$ and  $T^{ijl}$ are
symmetric, rank $1$, $2$, $3$ euclidean tensors respectively.
Once the integration over $d^2\bkt$ is performed, the $T^{i}$, $T^{ij}$ and
$T^{ijl}$ can only depend on the observable quantities $P_T$ and $\phi_h$,
{\it i.e.} the measured modulus and azimuthal phase of the final observed
hadron transverse momentum $\bfP _T$. Therefore, in a completely general way,
it must be
\bea
T^i     &=& P_T ^i \, S_1(P_T) \\
T^{ij}  &=& P_T ^i \, P_T ^j \, S_2(P_T) +\delta ^{ij} \, S_3(P_T) \\
T^{ijl} &=& P_T ^i \, P_T ^j \, P_T ^k \, S_4(P_T) +
(P_T ^i \, \delta ^{jl} + P_T ^j \, \delta ^{il} + P_T ^l \, \delta ^{ij})
\, S_5(P_T)\,,
\eea
where the $\bfP _T$ components ($P_T ^X = P_T \cos\phi_h$, $P_T ^Y =
P_T \sin\phi_h$) give the proper tensorial structure, while $S_1$--$S_5$ are
five scalar functions which can only depend on $P_T$ (modulus), and can easily
be determined by contracting Eqs.~(\ref{Ti})--(\ref{Tijl}) with some symmetric
tensorial structures ($P_T ^i$, $\delta ^{ij}$, etc..., as appropriate) to
obtain simple scalar relations. Finally, one finds
\bea
&&S_1(P_T) = \frac{1}{P_T} \int  d^2\bkt \; (\bkt \cdot \hat{\bfP} _T)
\, \Delta f (x,\kt) \, \Delta D (z,\pp) \label{S1}\\
&&S_2(P_T) = \frac{1}{P_T^2} \int d^2\bkt \; [2(\bkt \cdot \hat{\bfP} _T)^2
- \kt^2] \, \Delta f (x,\kt) \, \Delta D (z,\pp) \label{S2}\\
&&S_3(P_T) =  \int  d^2\bkt \; [\kt^2 - (\bkt \cdot \hat{\bfP} _T)^2] \,
\Delta f (x,\kt) \, \Delta D (z,\pp) \label{S3}\\
&&S_4(P_T) = \frac{1}{P_T^3} \int  d^2\bkt \; [4(\bkt \cdot \hat{\bfP} _T)^3
- 3\kt^2(\bkt \cdot \hat{\bfP} _T)] \, \Delta f (x,\kt) \, \Delta D (z,\pp)
\label{S4}\\
&&S_5(P_T) = \frac{1}{P_T} \int  d^2\bkt \; [\kt^2(\bkt \cdot \hat{\bfP}_T) -
(\bkt \cdot \hat{\bfP} _T)^3] \, \Delta f (x,\kt) \, \Delta D (z,\pp) \>.
\label{S5}
\eea
As a consequence, we have
\bea
&&\int d^2\bkt \; \cos\phi_\perp \, \Delta f \, \Delta D  =
\cos\phi_h \,\int d^2\bkt \, (\hat{\bfk}_\perp \cdot \hat{\bfP} _T)\;
\Delta f \, \Delta D \label{D12} \\
&&\int d^2\bkt \; \sin\phi_\perp \, \Delta f \, \Delta D  =
\sin\phi_h \,\int d^2\bkt \, (\hat{\bfk}_\perp \cdot \hat{\bfP} _T) \;
\Delta f \, \Delta D \label{D13} \\
&&\int d^2\bkt \; \cos ^2 \phi_\perp \, \Delta f \, \Delta D  =
\frac{1}{2} \int d^2\bkt \, \Big\{ 1+\cos 2\phi_h \,
[ 2(\hat{\bfk}_\perp \cdot \hat{\bfP} _T)^2 -1 ] \Big\}
\;\Delta f \, \Delta D \label{D14} \\
&&\int d^2\bkt \; \sin ^2 \phi_\perp \, \Delta f \, \Delta D  =
\frac{1}{2} \int d^2\bkt \, \Big\{ 1 - \cos 2\phi_h \,
[ 2(\hat{\bfk}_\perp \cdot \hat{\bfP} _T)^2 +1 ] \Big\}
\; \Delta f \, \Delta D \label{D15} \\
&&\int d^2\bkt \; \cos\phi_\perp \,\sin\phi_\perp \, \Delta f \, \Delta D  =
\cos\phi_h \, \sin\phi_h \int d^2\bkt \,
[ 2(\hat{\bfk}_\perp \cdot \hat{\bfP} _T)^2 - 1 ]
\; \Delta f \, \Delta D \label{D16} \\
&&\int d^2\bkt \; \cos ^3 \phi_\perp \, \Delta f \, \Delta D  =
\cos ^3 \phi_h \int d^2\bkt \, [4(\hat{\bfk}_\perp \cdot \hat{\bfP}_T)^3 -
3(\hat{\bfk}_\perp \cdot \hat{\bfP}_T)] \; \Delta f \, \Delta D \nonumber \\
&& \hspace{3.8cm}
+ \> 3 \cos\phi_h \int d^2\bkt \, [(\hat{\bfk}_\perp \cdot \hat{\bfP}_T) -
(\hat{\bfk}_\perp \cdot \hat{\bfP}_T)^3] \; \Delta f \, \Delta D \label{D17} \\
&&\int d^2\bkt \; \sin ^3 \phi_\perp \, \Delta f \, \Delta D  =
\sin ^3 \phi_h \int d^2\bkt \, [4(\hat{\bfk}_\perp \cdot \hat{\bfP}_T)^3 -
3(\hat{\bfk}_\perp \cdot \hat{\bfP}_T)] \; \Delta f \, \Delta D \nonumber \\
&& \hspace{3.8cm}
+ \> 3 \sin\phi_h \int d^2\bkt \, [(\hat{\bfk}_\perp \cdot \hat{\bfP}_T) -
(\hat{\bfk}_\perp \cdot \hat{\bfP}_T)^3] \; \Delta f \, \Delta D \label{D18} \\
&&\int d^2\bkt \; \cos ^2 \phi_\perp \sin\phi_\perp \, \Delta f \, \Delta D =
\cos ^2 \phi_h \sin\phi_h \int d^2\bkt \,
[4(\hat{\bfk}_\perp \cdot \hat{\bfP}_T)^3 -
3(\hat{\bfk}_\perp \cdot \hat{\bfP}_T)] \; \Delta f \, \Delta D \nonumber \\
&& \hspace{4.8cm}
+ \> \sin\phi_h \int d^2\bkt \, [(\hat{\bfk}_\perp \cdot \hat{\bfP}_T) -
(\hat{\bfk}_\perp \cdot \hat{\bfP}_T)^3] \; \Delta f \, \Delta D \label{D19} \\
&&\int d^2\bkt \; \cos\phi_\perp \sin ^2 \phi_\perp \, \Delta f \, \Delta D =
\cos\phi_h \sin ^2 \phi_h \int d^2\bkt \,
[4(\hat{\bfk}_\perp \cdot \hat{\bfP}_T)^3 -
3(\hat{\bfk}_\perp \cdot \hat{\bfP}_T)] \; \Delta f \, \Delta D \nonumber \\
&& \hspace{4.8cm}
+ \> \cos\phi_h \int d^2\bkt \, [(\hat{\bfk}_\perp \cdot \hat{\bfP}_T) -
(\hat{\bfk}_\perp \cdot \hat{\bfP}_T)^3] \; \Delta f \, \Delta D \>. \label{D20}
\eea
From these equations one can easily reconstruct
\bea
&&\int d^2\bkt \; \cos2\phi_\perp \, \Delta f \, \Delta D  =
\cos2\phi_h \,\int d^2\bkt \, [ 2(\hat{\bfk}_\perp \cdot \hat{\bfP} _T)^2 -1 ]\;
\Delta f \, \Delta D \label{D21} \\
&&\int d^2\bkt \; \sin2\phi_\perp \, \Delta f \, \Delta D  =
\sin2\phi_h \,\int d^2\bkt \, [ 2(\hat{\bfk}_\perp \cdot \hat{\bfP} _T)^2 -1 ] \;
\Delta f \, \Delta D \label{D22} \\
&&\int d^2\bkt \; \cos  3\phi_\perp \, \Delta f \, \Delta D  =
\cos 3 \phi_h \int d^2\bkt \, [4(\hat{\bfk}_\perp \cdot \hat{\bfP}_T)^3 -
3(\hat{\bfk}_\perp \cdot \hat{\bfP}_T)] \; \Delta f \, \Delta D \label{D23} \\
&&\int d^2\bkt \; \sin  3\phi_\perp \, \Delta f \, \Delta D  =
\sin 3 \phi_h \int d^2\bkt \, [4(\hat{\bfk}_\perp \cdot \hat{\bfP}_T)^3 -
3(\hat{\bfk}_\perp \cdot \hat{\bfP}_T)] \; \Delta f \, \Delta D \label{D24}\,.
\eea
All of these terms are easily recognizable in
Eqs.~(\ref{FUU})-(\ref{FUTsin(phi-2phiS)}).

\newpage
 
\section{Integration by rotation in the hadronic plane}\label{Int-p}

Eqs.~(\ref{FUU})-(\ref{FUTsin(phi-2phiS)}) can also be obtained in a simple 
way looking to a slightly different reference frame. Let us define
the production plane as the plane containing the virtual photon $\gamma^*$,
the proton momentum and the produced hadron $h$. We can define a new
$\gamma^*-p$ c.m. frame where the $X^{\prime}-Z^{\prime}$ plane is the
production plane. This new frame is rotated by an angle $\phi_{h}$ with 
respect to the c.m. frame $(\bfX, \bfY, \bfZ)$ depicted in Fig. 1 (we drop for 
simplicity the subscript $cm$):
\begin{eqnarray}
\hat{\bfX} &=& \hat{\bfX}^{\prime} \cos\phi_h -
             \hat{\bfY}^{\prime} \sin\phi_h \label{rotation1} \\
\hat{\bfY} &=& \hat{\bfX}^{\prime} \sin\phi_h +
             \hat{\bfY}^{\prime} \cos\phi_h \>. \label{rotation2}
\end{eqnarray}
Notice that $\hat{\bfX}^{\prime} = \hat{\bfP}_T = \hat{\bfh}$.
Any integration in Eqs.~(\ref{FUU})-(\ref{FUTsin(phi-2phiS)}), at fixed
values of the external variables, can be recast as the sum of one or more contributions of this kind:
\bea
&&\int  d^2\bkt \, k_{\perp} \cos\phi_\perp \, f(\kt, \pp) \;\hspace{1.5cm}
\int  d^2\bkt \, k_{\perp} \sin\phi_\perp  \, f(\kt, \pp)\;
\label{D3} \\
&&\int  d^2\bkt \, k_{\perp}^2 \cos2\phi_\perp  \, f(\kt, \pp) \;\hspace{1.5cm}
\int  d^2\bkt \, k_{\perp}^2 \sin2\phi_\perp  \, f(\kt, \pp) \;\hspace{1.5cm}
\label{D4} \\
&&\int  d^2\bkt \, k_{\perp}^3 \cos3\phi_\perp  \, f(\kt, \pp) \;\hspace{1.5cm}
\int  d^2\bkt \, k_{\perp}^3 \sin3\phi_\perp  \, f(\kt, \pp)
\label{D5} \>,
\eea
where 
\begin{equation}
\pp^2 = P_T^2 + z_h^2 \, \kt^2 - 2\, z_h 
(\bkt \cdot \bfP_T)\label{pperp-apxe}\,.
\end{equation}
Let us consider, for instance, Eq.~(\ref{D3}); using Eq.~(\ref{rotation1}), 
we have
\begin{eqnarray}
\int  d^2\bkt \, k_{\perp} \cos\phi_\perp \, f(\kt, \pp) &=&
\int  d^2\bkt \, k_{\perp}^X \, f(\kt, \pp)
= \int  d^2\bkt \, (\bkt \cdot \hat{\bfX}) \, f(\kt, \pp) \nonumber\\
&=&\int  d^2\bkt \left[ (\bkt \cdot \hat{\bfX}^{\prime}) \cos\phi_h
- (\bkt \cdot \hat{\bfY}^{\prime})\sin\phi_h \right] f(\kt, \bkt \cdot \hat{\bfX}^{\prime}) \label{parity-bis} \\
&=& \cos\phi_h \int d^2\bkt \, (\bkt \cdot \hat{\bfP}_T) \, f(\kt, \pp)\,,
\label{parity}
\end{eqnarray}
where in the step (\ref{parity-bis}) we have underlined that $f$ is a 
function of $(\bkt \cdot \hat{\bfP}_T)\equiv(\bkt \cdot \hat{\bfX}^{\prime})$ 
by means of Eq.~(\ref{pperp-apxe}).
With similar arguments we have, for all integrals of the kind
(\ref{D3})--(\ref{D5}):
\begin{eqnarray}
&&
\int d^2\bkt \, k_{\perp} \cos\phi_\perp 
\Rightarrow \cos\phi_h  \int d^2\bkt \,(\bfk_\perp \cdot \hat{\bfP}_T) \,
\\
&&
\int d^2\bkt \, k_{\perp} \sin\phi_\perp
\Rightarrow \sin\phi_h  \int d^2\bkt \,
(\bfk_\perp \cdot \hat{\bfP}_T)  \,
\\
&&
\int d^2\bkt \, k_{\perp}^2 \cos2\phi_\perp 
\Rightarrow \cos 2\phi_h\int d^2\bkt \,
[2 (\bfk_{\perp}\cdot \hat{\bfP}_T)^2 - k_{\perp}^2]  \,
\\
&&
\int d^2\bkt \, k_{\perp}^2 \sin2\phi_\perp 
\Rightarrow \sin2\phi_h \int d^2\bkt \,
[2 (\bfk_{\perp}\cdot \hat{\bfP}_T)^2 - k_{\perp}^2]  \,
\\
&&
\int d^2\bkt \, k_{\perp}^3 \cos3\phi_\perp 
\Rightarrow \cos 3\phi_h \int d^2\bkt \,
(\bfk_{\perp}\cdot \hat{\bfP}_T)\,
[4 (\bfk_{\perp}\cdot \hat{\bfP}_T)^2-3 k_{\perp}^2] \,
\\
&&
\int d^2\bkt \, k_{\perp}^3 \sin 3\phi_\perp
\Rightarrow \sin3\phi_h \int d^2\bkt \,
(\bfk_{\perp}\cdot \hat{\bfP}_T)\,
[4 (\bfk_{\perp}\cdot \hat{\bfP}_T)^2-3 k_{\perp}^2] \,,
\end{eqnarray}
which coincide with Eqs.~(\ref{D12}), (\ref{D13}) and (\ref{D21})--(\ref{D24}).

\newpage

\bibliographystyle{h-physrev5}

\bibliography{sample}
\end{document}